\documentstyle[version2,aps,preprint]{revtex}
\draft
\begin{document}
\preprint{To appear in Phys. Rev. B}

\begin{title}
Metallic Non-Fermi Liquid Phases\\
of an Extended Hubbard Model in Infinite Dimensions
\end{title}

\author{Qimiao Si and Gabriel Kotliar}

\begin{instit}
Serin Physics Laboratory, Rutgers University,
Piscataway, NJ 08855-0849, USA
\end{instit}

\begin{abstract}

We study an extended Hubbard model in  the limit of infinite
dimensions and its zero dimensional counterpart, a generalized
asymmetric Anderson model. In the impurity model we find three kinds
of mixed valence states: a) the usual strong coupling state in which a
resonance forms at the Fermi level; b) a weak coupling state in which
neither the impurity spin nor the impurity charge degrees of freedom
is quenched; and c) an intermediate coupling state where the spin but
not the charge degree of freedom is quenched. The corresponding phases
of the extended Hubbard model in infinite dimensions are a) a Fermi
liquid; b) a non-Fermi liquid state with incoherent charge and spin
excitations; and  c) a non-Fermi liquid state  with incoherent charge
but coherent spin excitations. The non-Fermi liquid phases are
incoherent metallic states  with vanishing quasiparticle residue,
self-similar local correlation functions, and asymptotically decoupled
charge and spin excitations. The non-Fermi liquid phases  occur for a
wide range of parameters.

\end{abstract}


\newpage

\newpage
\section{\bf Introduction}
\label{sec:intro}

Various anomalies in the normal state of the high $\rm T_c$ copper
oxides have motivated the current theoretical efforts to search for
models which exhibit metallic non-Fermi liquid
states.\cite{Anderson,RV} The mechanism for the breakdown of Fermi
liquid theory in one dimension is well understood. For weakly
interacting one dimensional fermion systems, the renormalization group
leads to the g-ology classification of spatially homogeneous metallic
states. The possible states are Luttinger liquids or those with
divergent CDW, SDW or superconducting correlation
functions.\cite{Solyom,Haldane1D} Bethe Ansatz, bosonization
and exact diagonalization methods have been used to show that this
classification persists for strong interactions provided that phase
separation does not intervene.\cite{giamarchi}

In higher than one dimensions, perturbative renormalization group
analysis has shown that, Fermi liquid theory does describe weakly
interacting fermion systems with a regular density of
states.\cite{Shankar} Therefore, the mechanism for the breakdown of
Fermi liquid theory is necessarily non-perturbative in the interaction
strength. Most of the previous approaches to this problem address
under what circumstances the residual interactions between the
quasiparticles can acquire singularities. Here we take an alternative
approach. We analyze the competition between local correlation and
itinerant effects without assuming quasiparticles to begin with. The
effect of interactions is treated non-perturbatively. This is carried
out in a controlled fashion, by studying this  problem in the limit of
infinite dimensionality.\cite{Vollhardt}

In the limit of infinite dimensions, all correlation functions of a
lattice fermion model can be reconstructed from the knowledge of the
{\it local} correlators, which can in turn be determined from those of
an {\it associated impurity model}.\cite{GK} The impurity model
describes some local degrees of freedom interacting with a
self-consistent electron bath. This limit
corresponds to a {\it dynamical} mean field theory, within which
non-trivial phases can be studied. The fact that only the low
frequency behavior of the correlators of the bath governs the low
frequency asymptotics of the correlators of the impurity model,
suggests that a
complete classification of the translationally invariant phases of
fermions in the limit of large dimensions could be within reach. The
Hubbard model with a Lorentzian density of states maps onto an
Anderson model with an {\it infinite} bandwidth, proving that this
model in large dimensions is a  Fermi liquid.\cite{GK}  The Hubbard
model in large dimensions with a {\it bounded} density of states, at
half filling and for large U, maps onto an Anderson impurity embedded
in an insulating medium\cite{Rozenberg,Krauth,Jarrell}, demonstrating
that at large U a frustrated Hubbard model is a Mott insulator. The
characterization of the state that results from doping this Mott
insulator is still  an open problem.

In this paper, we construct and study various  metallic non-Fermi
liquid states. These phases are  realized in  an extended (two band)
Hubbard model which contains the most general local density-density
and spin-spin interactions and a Lorentzian  density of states.
We stress however that our low energy analysis of the non-Fermi liquid
phases presented here is
more general than the particular extended Hubbard model that we
studied. We envision that many realistic Hamiltonians will be
described, at low energies, by the impurity actions that we study in
this paper.
The spin and charge  couplings  we discuss
will be generated dynamically in the process
of renormalization.

The impurity model associated with the extended Hubbard model is
a generalized Anderson model. It will be shown that the fact that the
impurity model is associated with a lattice model forces the impurity
to be near criticality. Qualitatively, this can be understood very
simply in the context of a weakly doped large U Hubbard model. In this
case the correlated site describes the local physics of the one band
Hubbard model and the bath describe the effects of the electrons near
that site. The self consistency condition restores translational
invariance and forces the correlated level to be near the chemical
potential. In a two band model describing a heavy (correlated) level
and a light conduction band  the heavy level participates in the
charge transport only when it is situated
near the chemical potential of the conduction band.
The low energy description  of the local physics of the one and the
two band models is very similar. The role of the light conduction band
in the two band model is played by the incoherent part of the one
particle Green's function in the one band model.

To study the low energy behavior of the correlators of the impurity
model near the critical point, we set up a renormalization group
analysis for this problem. Our strategy is to set up an expansion in
terms of the hopping amplitudes between the local configurations of
the impurity problem. This is an extension of the classic work of
Haldane.\cite{Haldane} The resulting expression of the partition
function resembles that of a classical Coulomb gas. The
hopping amplitudes correspond to fugacities of the Coulomb gas
representation. The fugacities are proportional to the amplitude for
making a transition between two different atomic states. When the
amplitude for making transitions between states of different charge
grows as we go towards longer time scales the system is a Fermi
liquid. When this amplitude  renormalizes to zero, quantum coherence
is destroyed and Fermi liquid theory breaks down. We cast the
breakdown of  Fermi liquid theory in the framework of the macroscopic
quantum tunneling (MQT) problem.\cite{Leggett} The transitions between
Fermi liquid and non-Fermi liquid phases are extensions of the well
known localization transitions in the MQT problem with one essential
difference (responsible for a richer phase diagram) that we deal with
a special {\it three level} system describing the local spin and
charge degrees of freedom instead of the canonical {\it two level}
system studied in the MQT literature.

The possible phases are determined from the possible fixed points of
the RG. We carry out the analysis of a more general SU(N) impurity
model. The case N=2 corresponds to the three level system (empty site
and SU(2) spin). We identify, besides the strong coupling phase in
which all excitations are coherent and the weak coupling phase in
which all excitations are incoherent, a new class of metallic state of
matter in which spin excitations are coherent while charge excitations
are incoherent. We call this an intermediate coupling phase. Both the
intermediate coupling and weak coupling phases correspond to
incoherent metallic states with vanishing quasiparticle residue,
self-similar local correlation functions, and asymptotically decoupled
charge and spin excitations. The low energy behaviors are not
described by Fermi liquid theory.

The N=1 case reduces to the well studied resonant level
model.\cite{Wieg,Schlot} It is well known that orthogonality effects
are more pronounced when there are bound states which can violate the
unitarity bounds of the coupling constants or in the presence of many
channels of conduction electrons.\cite{Hakim} This later point is
emphasized by Ruckenstein and Varma in the search of a mechanism for
Marginal Fermi liquid states in realistic models. Along this line,
Giamarchi {\it et al.} have recently suggested  that an impurity model
containing a bath with {\it several channels} of screening electrons
is closely related to the copper oxides.\cite{Giamarchi} More recently
Perakis {\it et al.} suggested that the presence of several channels
of screening electrons can give rise to non-Fermi liquid behavior
similar to the one found in the multichannel Kondo
problem.\cite{Perakis} The addition of extra channels of conduction
electrons\cite{HaldaneS} in our formalism amounts to a change in the
initial conditions of the renormalization group flow.

The  paper is organized as follows. In Section \ref{sec:ehm1}, we
introduce the extended Hubbard model and derive, in the limit of
infinite dimensions, the associated impurity problem in a
self-consistent medium. This impurity problem is a generalized
asymmetric Anderson model. The qualitative physics associated with
such an impurity problem is discussed in Section \ref{sec:ehm2}.
The justification for the associated impurity problem being near
criticality is given in Section \ref{sec:ehm3}. In Section
\ref{sec:parfun1}, we construct the atomic expansion for the partition
function of the impurity problem. The resultant partition function is
a summation over histories of the local degrees of freedom. It is
shown in Section \ref{sec:parfun2} that, this partition function
corresponds to a particular realization of a class of one dimensional
classical spin models with long range interactions considered by
Cardy,\cite{Cardy} extended to account for a symmetry breaking field.
Using Cardy's procedure, we derive in Section \ref{sec:flow} the
renormalization group equations, and analyze the resulting
universality classes and crossover behavior. In section \ref{sec:phases}
we  establish the nature of the Fermi liquid and
the  non-Fermi liquid phases of the extended Hubbard model. A summary of
the phase diagram is given in Section \ref{sec:phases3}. In Section
\ref{sec:phases4}, we discuss the relevance of our results to more
realistic  densities of states. In section \ref{sec:conclu}
we conclude with a discussion of the relevance of our results to
realistic finite dimensional systems. The paper  includes three
appendices. Appendix \ref{sec:bos} summarizes the bosonization
procedure. Appendix \ref{sec:scaling.h} gives the details of the
derivation of the renormalization group equations. Finally, in
Appendix \ref{sec:corr}, we derive the correlation functions in the
weak coupling mixed valence regime. A brief account of the
renormalization group flow and its implication for the extended
Hubbard model in infinite dimensions has been given in Ref.
\cite{short}.

\bigskip
\section{\bf The Extended Hubbard Model in Infinite Dimensions}
\label{sec:ehm}

\subsection{ The Extended Hubbard Model in Infinite Dimensions\\
and its Associated Impurity Model}
\label{sec:ehm1}

We study the following extended Hubbard model

\begin{eqnarray}
H = &&\sum_{ij} \sum_{\sigma=1}^N (t_{ij}-\mu\delta_{ij})
c^{\dagger}_{i\sigma} c_{j\sigma} + \sum_i \sum_{\sigma=1}^N (
{\epsilon}^o_d -\mu) d^{\dagger}_{i\sigma} d_{i\sigma} + {U \over 2} \sum_i
\sum_{\sigma \ne \sigma'} d^{\dagger}_{i\sigma} d_{i\sigma }
d^{\dagger}_{i\sigma ' } d_{i\sigma ' }\nonumber\\
&&+\sum_i \sum_{\sigma=1}^N t ( d^{\dagger}_{i\sigma}
c_{i\sigma} + h.c. ) + {V_1 \over N} \sum_i
\sum_{\sigma,\sigma ' } c^{\dagger}_{i\sigma}
c_{i\sigma} d^{\dagger}_{i\sigma ' } d_{i\sigma '}
+ {V_2 \over N} \sum_i \sum_{\sigma,\sigma ' }
c^{\dagger}_{i\sigma} c_{i\sigma ' }
d^{\dagger}_{i\sigma ' } d_{i\sigma }
\label{hamiltonian.ehm}
\end{eqnarray}
It includes two species of electrons. The uncorrelated conduction $c$
electrons have the hopping matrix $t_{ij}$ or, equivalently, the
dispersion $\epsilon_k$. The localized $d$ electrons have an energy
level ${\epsilon}^o_d$, and an on-site Coulomb interaction $U$ which
is taken to be infinity. These two species of electrons are coupled,
at every site, through a hybridization $t$, a density density
interaction ${V_1 \over N}$, and a spin exchange interaction ${V_2
\over N}$. For generality, we work with arbitrary spin degeneracy $N$,
which will also allow us to make contact with results from large $N$
expansion. Finally, the chemical potential is $\mu$.

In the spinless case (N=1) this model is reduced to a generalized
Falicov-Kimball model which includes, besides the usual interaction
$V_1$, a hybridization $t$ term. In the limit of vanishing
hybridization, the Falicov-Kimball model is exactly soluble in
infinite dimensions.\cite{Brandt,Janis} When the density of states is
a Lorentzian \cite{lorentzian,GK} all the correlation functions can be
computed analytically and the origin of the breakdown of Fermi liquid
theory in the model can be understood in simple physical
terms.\cite{fk} There is a range of electron densities where the
chemical potential coincides with the position of the renormalized d
electron level. We referred to that situation as the pinning of the
chemical potential. We notice however that the Falicov-Kimball model
should be understood as  the zero hybridization  limit  of the model
(\ref{hamiltonian.ehm}).\cite{fk} In this paper we show that there are
regions of parameter space in which the hybridization dynamically
renormalizes to zero at low energies and establish that the
Falicov-Kimball fixed point in the pinned region has a finite basin of
attraction.

The  spin ${1\over 2}$ case (N=2)  is a version of the extended
Hubbard model first studied in the context of the mixed valence
problem.\cite{MV} It has some similarities with the extended
Hubbard model relevant to the high ${\rm T_c}$ systems proposed by
Emery\cite{Emery} and Varma, Schmitt-Rink, and Abrahams.\cite{VSA}

The infinite dimension limit of a lattice fermion model is defined
through an appropriate scaling of the hopping term,\cite{Vollhardt}
which in our case is that of the conduction electrons. It has been
shown that, one can naturally associate with each  lattice model in
infinite dimensions an impurity model. This impurity model is obtained
by integrating out all the degrees of freedom except for those living
at the origin.\cite{GK,GKS}

Following this procedure we obtain the single site effective action of
our extended Hubbard model:

\begin{eqnarray}
S_{site} (G_o)= &&-\int^{\beta}_0 d\tau \int^{\beta}_0 d\tau '
\psi^\dagger (\tau )G^{-1}_o (\tau - \tau' ) \psi ( \tau' )\nonumber\\
&&+ \int^{\beta}_0 d\tau ( {V_1 \over N} \sum_{\sigma,\sigma ' }
c^{\dagger}_{\sigma} c_{\sigma} d^{\dagger}_{\sigma ' } d_{\sigma '} +
{V_2 \over N} \sum_{\sigma,\sigma ' } c^{\dagger}_{\sigma} c_{\sigma '
} d^{\dagger}_{\sigma ' } d_{\sigma } + {U \over 2} \sum_{\sigma \ne
\sigma'} d^{\dagger}_{\sigma} d_{\sigma } d^{\dagger}_{\sigma ' } d_{\sigma ' }
)
\label{siteaction}
\end{eqnarray}
where $\psi^\dagger = ( c^\dagger , d^\dagger )$. The Weiss effective
field $G^{-1}_o$ is a $2 \times 2$ matrix. It describes the effect of
all the Feynman trajectories in which the electron at site zero leaves
the origin, explores the lattice, and returns to the origin.
Translational invariance demands that all the local correlation
functions of the lattice model  are the same as the  correlation
functions of the impurity model leading  to the following
self-consistency equations:

\begin{eqnarray}
\int^{\infty}_{-\infty} d\epsilon &&
N_o(\epsilon)
\left( \begin{array}{ll}
        i\omega_{n} +\mu-\epsilon-\Sigma_{cc}^{imp}~ ~ ~ ~ ~ ~ ~ ~ ~ &
t- \Sigma_{cd}^{imp}\\
        t-\Sigma_{dc}^{imp}~ ~ ~ ~ ~ ~ ~ ~ ~ & i\omega_{n} +\mu
-{\epsilon}^o_d -\Sigma_{dd}^{imp} \end{array}
\right)^{-1}\nonumber\\
&&=~
\left( \begin{array}{ll}
        (G_o^{-1})_{cc}-\Sigma_{cc}^{imp}~ ~ ~ ~ ~ ~ ~ ~ ~ &
(G_o^{-1})_{cd}- \Sigma_{cd}^{imp}\\
        (G_o^{-1})_{dc}-\Sigma_{dc}^{imp}~ ~ ~ ~ ~ ~ ~ ~ ~ &
(G_o^{-1})_{dd}-\Sigma_{dd}^{imp} \end{array}
\right)^{-1}
\label{consist}
\end{eqnarray}
Here the non-interacting density of states, $N_0(\epsilon)= \sum_k
\delta(\epsilon -\epsilon_k)$ where $\epsilon_k$ the conduction
electron dispersion, is the only place where the nature of the lattice
enters.  $\Sigma^{imp}$ is a functional of $G_o$ defined by the
self-energy matrix of the single-site impurity model defined in Eq.
(\ref{siteaction}).  After solving the system of equations (2) and (3)
for $G_o$ we can evaluate  the self-energy of the lattice model
$\Sigma =\Sigma_{imp}( G_o)$.

For a Lorentzian density of states, $G_o$ can be solved
explicitly\cite{GK,fk}:

\begin{eqnarray}
\left( \begin{array}{ll}
        (G_o^{-1})_{cc}~ ~ ~ ~ ~ ~ ~ ~ ~ & (G_o^{-1})_{cd}\\
(G_o^{-1})_{dc}~ ~ ~ ~ ~ ~ ~ ~ ~ & (G_o^{-1})_{dd} \end{array}
\right)~=~
\left( \begin{array}{ll}
        i\omega_n + \mu + i\Gamma sgn\omega_n~ ~ ~ ~ ~ ~ ~ ~ ~ & t\\
t ~ ~ ~ ~ ~ ~ ~ ~ ~ & i\omega_n+\mu-{\epsilon}_d^o \end{array}
\right)
\label{GoLoren}
\end{eqnarray}
where $\Gamma$ is the width of the Lorentzian.  A more general density
of states, however does not alter the qualitative results derived
here, provided that the system is metallic and the hybridization $t$
is smaller than the bandwidth of the conduction electrons. This will
be discussed further in Section \ref{sec:phases4}.

Our single site action describes an impurity model :

\begin{eqnarray}
S_{imp}= &&\int^{\beta}_0 d\tau \sum_{k\sigma}c_{k\sigma}^\dagger
(\partial_{\tau}+\mu- \epsilon_k)c_{k\sigma} +d_{\sigma}^\dagger
(\partial_{\tau}+\mu-{\epsilon}^o_d)d_{\sigma} +\sum_{\sigma} t (
d^{\dagger}_{\sigma} c_{\sigma} + h.c. )\nonumber\\
&&+{V_1 \over N} \sum_{\sigma,\sigma ' } c^{\dagger}_{\sigma}
c_{\sigma} d^{\dagger}_{\sigma ' } d_{\sigma '} + {V_2 \over N}
\sum_{\sigma,\sigma ' } c^{\dagger}_{\sigma} c_{\sigma ' }
d^{\dagger}_{\sigma ' } d_{\sigma } + {U \over 2} \sum_{\sigma \ne
\sigma'} d^{\dagger}_{\sigma} d_{\sigma } d^{\dagger}_{\sigma ' }
d_{\sigma ' }
\label{impaction}
\end{eqnarray}
where an electron bath, with dispersion $\epsilon_k$, has been
introduced to generate the Weiss field $G_o^{-1}$ given in Eq.
(\ref{GoLoren}). This is the generalized Anderson impurity model.

\subsection{Qualitative Physics of the Generalized Anderson Model}
\label{sec:ehm2}

The Hamiltonian of the generalized Anderson impurity problem
corresponding to  the action in Eq. (\ref{impaction}) is  given by:

\begin{eqnarray}
H_{\rm imp}= &&\sum_{k\sigma} E_k c_{k\sigma}^\dagger c_{k\sigma}
+E^o_d d_{\sigma}^\dagger d_{\sigma} +\sum_{\sigma} t (
d^{\dagger}_{\sigma} c_{\sigma} + h.c. )\nonumber\\
&&+ {V_1 \over N} \sum_{\sigma,\sigma ' } c^{\dagger}_{\sigma}
c_{\sigma} d^{\dagger}_{\sigma ' } d_{\sigma '} + {V_2 \over N}
\sum_{\sigma,\sigma ' } c^{\dagger}_{\sigma} c_{\sigma ' }
d^{\dagger}_{\sigma ' } d_{\sigma } +{U \over 2} \sum_{\sigma \ne
\sigma'} d^{\dagger}_{\sigma} d_{\sigma } d^{\dagger}_{\sigma ' }
d_{\sigma ' }
\label{hamiltonian}
\end{eqnarray}
where the $d$ electron level $E_d^o$ is given by

\begin{eqnarray}
E_d^o = {\epsilon}_d^o-\mu
\label{dlevel}
\end{eqnarray}
and the bath electron dispersion $E_k=\epsilon_k-\mu$.

In this subsection, we give a qualitative discussion of the physics
associated with this model. We separate the discussion for the
spinless and spinful cases.

In the spinless case (N=1) this model is reduced to the resonant level
model. It was previously established\cite{Wieg,Schlot} that, the
resonant level model can be mapped onto an anisotropic spin ${1\over
2}$ Kondo problem through the following: a) the hybridization $t$ is
identified with the transverse Kondo exchange ${1 \over 2} J_{\perp}$;
b) the density-density interaction $V_1$, when written in a
particle-hole symmetric form, is related to the longitudinal component
of the Kondo exchange $J_{\parallel}$ through
identifying the corresponding phase shifts: $(1-2\delta/\pi)^2
\rightarrow {2}(1-2 \delta_{\parallel}/\pi )^2$, where $\delta
=tan^{-1}(\pi \rho_oV_1/2)$ and $ \delta_{\parallel} =
tan^{-1}(\pi \rho_o J_{\parallel}/4) $, with $\rho_o$ the conduction
electron density of states at the Fermi level; and c) the
$d$ electron level $E_d^o$ is mapped onto the magnetic field $(-H)$
applied on the local moment. Such a mapping allows a parallel
discussion of the qualitative physics of the resonant level model and
that of the Kondo problem.

We recall that, at zero magnetic field, the Kondo problem has weak
coupling fixed points when the exchange coupling is
ferromagnetic: the transverse Kondo exchange is renormalized to zero.
For an antiferromagnetic exchange coupling, it flows to
a strong coupling fixed point.\cite{Yuval,Wilson} The transition
can be viewed as an unbinding of defects ({\it spin flips})
in the path integral representation fo the partition function to be
derived in section \ref{sec:parfun}, as in the Kosterlitz-Thouless
transition of the $2D$ $XY$ model.\cite{Kosterlitz} In complete
analogy, when the $d$ level is located at the chemical potential, the
resonant level model has a weak coupling fixed  line for a range of
attractive interactions given by $2\delta/\pi <
-(\sqrt{2}-1)$.\cite{Wieg,Schlot} The hybridization becomes relevant
for $2\delta/\pi >-(\sqrt{2}-1)$, at which a strong coupling phase
emerges at low energies. The zero-temperature transition between these
phases involves the unbinding of defects describing transitions
between states with different charge ({\it charge flips}) in the path
integral representation of the partition function to be described in
section \ref{sec:parfun}.

Applying a large magnetic field at the impurity in the Kondo problem
polarizes the impurity spin, cuts off the infrared singularities, and
therefore destroys the Kondo process. Similarly in the resonant level
model, when the $d$ level is far away from the chemical potential, the
$d$ orbital is either fully occupied or empty, and the resonant level
model becomes a potential scattering problem. Interesting physics
occurs in the mixed valence regime, when the renormalized $d$-level,
denoted as $E_d^*$, is close to zero.

The nature of these fixed points can also be discussed in parallel.
In the ferromagnetic Kondo problem in  zero magnetic field, the
spectrum is composed of conduction electrons and a local spin mode,
which are asymptotically decoupled at low energies. There is no energy
scale in the problem; Correlation functions have long time algebraic
behavior and do not have the Fermi liquid form. Similarly the spectrum
of the resonant level model in the weak coupling regime, when the
$d$-level is located at the Fermi energy of the conduction electrons,
is composed of the local $d$-electron and the conduction electrons
which are asymptotically decoupled.  The $d$ electron spectral
function, as well as local correlation functions, exhibit power law
behavior in frequency with interaction-dependent exponents, due to the
finite value of the renormalized density-density interaction $V_1^*$.
The divergence occurs at zero frequency when $E_d^*=0$. In both cases,
the low energy excitations are incoherent, and can not be analytically
continued to those of non-interacting electron systems.

In the antiferromagnetic Kondo problem, the local spin mode is
quenched below the  Kondo energy scale $T_K$ through the formation of
quasiparticles, which correspond to coherent spin excitations. In the
resonant level model at the strong coupling regime, the
renormalized hybridization $t^*$ between the local $d$ and conduction
electrons is finite. This renormalized hybridization sets the energy
scale, below which the local $d$ electron {\it retains} its
free-electron character and combines with
the conduction electrons to form quasiparticles.
These quasiparticles describe coherent charge excitations.
In both cases, the low energy behavior is described by a Fermi liquid.

The above discussion indicates that, whether or not the hybridization
is relevant determines the nature of the low energy behavior.
This point can be further seen by noticing that the wave function
renormalization factor associated with the $d$ electron is equivalent
to the renormalization of the hybridization: therefore, whether or not
the the renormalized hybridization is non-zero determines the
existence or vanishing of the quasiparticle residue. This motivates us
to study the flow of the hybridization,  along with other couplings.

For spinful case ($N \ge 2$) the generalized Anderson model has low
energy excitations in both the charge and spin channel.  We will show,
using the RG flow, that the interplay between charge and spin dynamics
lead to three kinds of mixed valence states, characterized by
different ways the hybridization and the exchange coupling are
renormalized.

{\it Strong coupling mixed valence fixed point.~} Both the
hybridization and exchange coupling are relevant. As in the
antiferromagnetic Kondo problem and in the strong coupling regime of
the resonant level model, the spectral functions at low energies are
described in terms of a resonance around the chemical potential. The
width of the resonance sets the scale for the low energy coherent
quasiparticle excitations. This is in the same universality class as
the usual strong coupling phase of the Anderson model.

{\it Weak coupling mixed valence fixed points.~} Both the
hybridization and the exchange coupling are irrelevant. The system is
self-similar; Correlation functions in the charge and spin
channels have algebraic behavior similar to those of the ferromagnetic
Kondo problem and of the weak coupling regime of the resonant level
model respectively. Both the charge and spin excitations are
incoherent.

{\it Intermediate coupling mixed valence fixed points.~} Here, the
hybridization is irrelevant, while the exchange coupling is relevant.
The system is self-similar in the charge channel, and exhibits a
finite energy scale in the spin channel. The correlation functions in
the charge channel have an algebraic behavior similar to those of the
weak coupling regime of the resonant level model, while the
correlation functions in the spin channel are characterized by a
resonance in analogy to those of the antiferromagnetic Kondo problem.
Therefore, the low energy charge excitations are incoherent while the
low energy spin excitations are coherent.

The intermediate mixed valence phase occurs because the
two kinds of defects (spin flips and charge flips) can unbind at
different stages.\cite{Nelson} We will establish that, due to the
special form of the coupling between the spin and charge dynamics,
a phase with coherent charge excitations and incoherent spin
excitations can not exist.

\subsection{Pinning of the Local States at Chemical Potential}
\label{sec:ehm3}

In this subsection we show that, the fact that the impurity model is
the result of a mapping from a lattice model forces the impurity to be
near criticality.

The local Green's functions of a lattice fermion model in infinite
dimensions are given by those of the impurity problem. Therefore,
the density of $d$ electrons plus the local density of  $c$ electrons in the
ground state of the impurity model Eq. (\ref{hamiltonian}) equals the
total electron density ($n$) of the original lattice problem Eq.
(\ref{hamiltonian.ehm}).

\begin{eqnarray}
\sum_{\sigma} <d_\sigma ^\dagger d_\sigma> + \sum_{\sigma}
<c_{0,\sigma}^\dagger c_{0,\sigma}> = n
\label{density}
\end{eqnarray}

To determine $n_d$ within the impurity model Eq. (\ref{hamiltonian}),
we first note that the $d$-level is renormalized from its bare value.
This happens due to two sources, a short time
Hartree-like renormalization due to the interactions and
a long time contribution due to the hybridization which
is described by the RG flow of
section \ref{sec:phases}.

To study the short time renormalization, we consider
the non-flipping part of the Hamiltonian given in
Appendix \ref{sec:bos}. The shift of the $d$-level $\Delta E_d$
comes from the
change of the ground state energy of the conduction electron sea, when
a $d$-electron is present.  The renormalized $d$-level is given by
\begin{eqnarray}
\tilde{E}_d (\mu ) ={\epsilon}_d^o -\mu + \Delta E_d (\mu)
\label{pin1}
\end{eqnarray}
Here $\Delta E_d$ depends on the chemical potential $\mu$ since
the shift of
the ground state energy changes as $\mu$ is varied.

There exists a critical chemical potential, $\mu=\mu_c^o$, at which
the renormalized $d$ level lies at the chemical potential,

\begin{eqnarray}
\tilde{E}_d ({\mu}_c^o ) = 0
\label{pin2}
\end{eqnarray}

In the absence of hybridization, such a condition indicates that, at
$\mu=\mu_c^o$, the heavy electron lies right at the Fermi level of the
electron bath. Both $n_d$ and $n_c$ jump as $\mu$ varies
through $\mu_c$. Therefore, {\it a finite range of electron densities}
correspond to the chemical potential $\mu_c^o$.\cite{fk} Over this range of the
electron densities, the $d$-level is pinned at the chemical potential.
Within this pinned density range, there is no energy barrier for the
transition from one local charge state to another.

The essential question is whether, when the hybridization term is
present, the chemical potential can still be adjusted to a value
$\mu_c$ so that the renormalized $d$-level is equal to zero. In the
language of critical phenomena, our procedure here is equivalent to
renormalizing a massless field theory. In Section \ref{sec:phases}, we
will show through the RG analysis that this can indeed be achieved
when the hybridization is irrelevant.

\bigskip
\section{Partition Function of the Generalized Anderson Model}
\label{sec:parfun}

We now turn to the RG analysis of the generalized
Anderson model Eq. (\ref{hamiltonian}) near the mixed valence regime
where the renormalized $d$ level is near the chemical potential.

To derive the RG flow, our strategy is to first construct an expansion
around the atomic limit, in terms of the hopping amplitudes between
the local atomic configurations. It leads to a partition function
written as a summation over histories associated with the local
degrees of freedom. This in turn is equivalent to a $0+1$ dimensional
statistical mechanical model with long range interactions between the
flipping events. Such an atomic expansion  originates in the work of
Anderson and Yuval in the Kondo problem\cite{Yuval} and of Haldane in
the asymmetric Anderson model\cite{Haldane}. Here we generalize it to
arbitrary local configurations, and construct a closed set of RG
equations. We stress that, this atomic expansion is perturbative in
the hoppings, but non-perturbative in the interactions and the
symmetry breaking fields.

To illustrate the methodology, we recall that, RG equations for
Kondo-like impurity problems have been derived using a) poor man's
scaling\cite{Yuval,Haldane}; b) multiplicative RG\cite{Fowler}; and c)
mapping to one-dimensional statistical mechanical
problem.\cite{Yuval,Haldane} In the most general form, both the poor
man's scaling and the multiplicative RG are difficult to carry out
systematically:  many couplings are present and it is not {\it a
priori} clear which combinations of these couplings come into the
flow. We will show that, for a general set of local configurations,
the partition functions derived from the atomic expansion can be
mapped onto a class of discrete classical spin chain models with a
long range $1/r^2$ interaction in symmetry breaking fields. This will
allow us to carry out a systematic RG analysis in the general case, by
generalizing Cardy's analysis of these models\cite{Cardy} to
incorporate the symmetry breaking fields.
We note that, Haldane's work\cite{Haldane} for the asymmetric Anderson
model emphasizes the importance of a symmetry breaking field (i.e. the
impurity level) as a consequence of the lack of particle-hole
symmetry. Since the focus is on identifying various regimes {\it
within} the strong coupling Fermi liquid phase, scaling equations are
constructed for the hybridization and the symmetry breaking field
only. In Cardy's work\cite{Cardy} for the general class of one
dimensional classical discrete spin chains with long range
interactions, RG flow is constructed for all coupling constants, but
without incorporating the symmetry breaking fields. The effect of the
symmetry breaking fields in this scheme has been studied previously in
other context.\cite{gabi} For our purposes of identifying different
low energy fixed points and the associated transitions/crossovers
between them, it is essential to study the renormalization of all the
coupling constants in the generalized Anderson model. Meanwhile,
particle-hole asymmetry generates non-zero effective fields even if we
start from zero values of the bare fields. By combinging the formalism
used in the works of Haldane and Cardy, we are able to derive the
systematic RG equations for the generalized Anderson model near
criticality, and identify and study Fermi liquid and non-Fermi liquid phases.

In this section, we give the details of the atomic expansion for the
partition function of the generalized Anderson model. In the next
section, we derive the RG flow and establish the nature of the
resulting fixed points.

\subsection{The Partition Function as a Summation over Histories}
\label{sec:parfun1}

For the generalized Anderson model at $U=\infty$,\cite{finiteU} the
local degrees of freedom are characterized by the $N+1$ $d$-electron
states, $|\alpha>=|0>$, or $|\sigma_m>$ for $m=1,...,N$.

To perform the atomic expansion, we first separate the impurity
Hamiltonian into two parts

\begin{eqnarray}
H ~=~H_0+ H_f
\label{separate}
\end{eqnarray}
where $H_0$ is diagonal in the space of the local states:

\begin{eqnarray}
H_0= &&\sum_{k~\sigma}E_k c_{k\sigma}^\dagger c_{k\sigma}
+E^o_d d_{\sigma}^\dagger d_{\sigma}
+ {U\over 2} \sum_{\sigma \ne \sigma'} d^{\dagger}_{\sigma} d_{\sigma }
d^{\dagger}_{\sigma ' } d_{\sigma ' }\nonumber\\
&&+ {V_1 \over N} \sum_{\sigma,\sigma ' } c^{\dagger}_{\sigma} c_{\sigma}
d^{\dagger}_{\sigma ' } d_{\sigma '}
+ {V_2^{\parallel} \over N} \sum_{\sigma } c^{\dagger}_{\sigma} c_{\sigma }
d^{\dagger}_{\sigma  } d_{\sigma }
\label{s0}
\end{eqnarray}

The remaining part

\begin{eqnarray}
H_f= \sum_{\sigma} t ( d^{\dagger}_{\sigma} c_{\sigma} + h.c. )
+ {V_2^{\perp} \over N} \sum_{\sigma , \sigma ' \ne \sigma}
d^{\dagger}_{\sigma  } d_{\sigma' } c^{\dagger}_{\sigma '} c_{\sigma }
\label{hf}
\end{eqnarray}
flips from one local state to another.

We can write $H_0$ in terms of the projection operators $X_{\alpha
\alpha}=|\alpha><\alpha|$,

\begin{eqnarray}
H_0 = \sum_{\alpha} H_{\alpha} X_{\alpha \alpha}
\label{s0.1}
\end{eqnarray}
where

\begin{eqnarray}
H_{\alpha} = E_{\alpha}+\sum_{\gamma} V_{\alpha}^{\gamma} c^\dagger_{\gamma}
c_{\gamma} + \sum_{k~\gamma} E_k c^\dagger_{k \gamma} c_{k \gamma}
\label{halpha}
\end{eqnarray}
Here $c_{\gamma}^\dagger$ creats a Wanniar state of the conduction
electrons of spin $\gamma$ at the impurity site. The local levels are
$E_0=0$, $E_{\sigma} = E_d^0$, and the potentials are given by
\begin{eqnarray}
&&V_{\sigma}^{\gamma} = {V_1 \over N}+ {V_2^{\parallel} \over N}
\delta_{\gamma \sigma}\nonumber\\
&&V_0^{\gamma}=0
\label{potential}
\end{eqnarray}

Meanwhile, the flipping part can be decomposed as follows,

\begin{eqnarray}
H_f= \sum_{\alpha \ne \beta} Q(\alpha,\beta)
\label{hf.2}
\end{eqnarray}
where

\begin{eqnarray}
Q(\alpha,\beta)
= |\alpha> <\alpha|H_f|\beta> <\beta|
\label{q}
\end{eqnarray}
describes the process flipping from the local state $|\beta>$ into
$|\alpha>$. Specifically,

\begin{eqnarray}
&&Q(\sigma,0)=Q^\dagger(0,\sigma)=td^\dagger_{\sigma}c_{\sigma}\nonumber\\
&&Q(\sigma,\sigma ')= {V_2^{\perp} \over N} d^\dagger_{\sigma}d_{\sigma '}
c^\dagger_{\sigma '}c_{\sigma} (1-\delta_{\sigma \sigma'})
\label{q2}
\end{eqnarray}

The partition function

\begin{eqnarray}
Z= \int Dc~D d ~ ~exp [-(S_0 + \int_0^{\beta}d\tau H_f(\tau) )]
\label{part}
\end{eqnarray}
can be expanded in $H_f$

\begin{eqnarray}
Z= \sum_{n=0}^{\infty}~ ~&&\int_{0}^{\beta}d\tau_n
...\int_{0}^{\tau_{i+1}} d \tau_i
...\int_{0}^{\tau_{2}} d \tau_1~
\sum_{\alpha}
A (\alpha; \tau_n,...,\tau_1)
\label{pert1}
\end{eqnarray}
where the transition amplitude, for a given initial (and final)
state $|\alpha >$ and a sequence of flipping times
$\tau_1<...<\tau_n$, is given by

\begin{eqnarray}
A (\alpha; \tau_n,...,\tau_1) = (-1)^n \int Dc~D d
<\alpha|T[ exp(-\beta H_0 )
H_f(\tau_n)...H_f(\tau_i)...H_f(\tau_1)] |\alpha>
\label{amplitude}
\end{eqnarray}

For each transition amplitude, the path integral over the $d$-states can
be carried out through inserting a complete set of local states at
every discrete imaginary time, leading to

\begin{eqnarray}
A (\alpha; \tau_n,&&...,\tau_1)= (-1)^n \sum_{\alpha_2, ..., \alpha_n}
\int Dc ~exp[-H_{\alpha} (\beta-\tau_{n})]
Q'(\alpha,\alpha_{n})...\times\nonumber\\
&& \times exp[-H_{\alpha_{i+1}} (\tau_{i+1}-\tau_{i})]
Q'(\alpha_{i+1},\alpha_{i})~
exp[-H_{\alpha_i} (\tau_i-\tau_{i-1})]... \times \nonumber\\
&&\times exp[-H_{\alpha_2} (\tau_2-\tau_1)]~ ~
Q'(\alpha_2,\alpha) ~exp[-H_{\alpha} \tau_1]
\label{pert5}
\end{eqnarray}
Here, $\alpha,\alpha_2,...,\alpha_n$ and $\tau_1,...,\tau_n$ label a
Feynman trajectory: the local configuration starts from
$|\alpha_1>=|\alpha>$, changes from $|\alpha_i>$ to $|\alpha_{i+1}>$
at time $\tau_i$, ($i=1,...,n-1$), and returns to
$|\alpha_{n+1}>=|\alpha>$ at $\tau_n$. This is illustrated in
Fig. \ref{hop}.

The flipping operator
$Q'(\alpha_{i+1},\alpha_{i})=<\alpha_{i+1}|H_f|\alpha_{i}>$ can be
separated as follows,

\begin{eqnarray}
<\alpha|H_f|\beta>
= y'_{\alpha\beta} O' (\alpha,\beta)
\label{q1}
\end{eqnarray}
Here $y_{\alpha \beta}'$ is the hopping amplitude associated with the
flipping event $(\alpha, \beta )$: $y'_{0,\sigma}=y'_{\sigma,0}$=t,
$y'_{\sigma,\sigma'}={V_2^{\perp} \over N}$. $O'(\alpha,\beta )$
is composed of conduction electron operators which describe the
distortion in the sea of conduction electrons associated with the
flipping event: $O'(\sigma,0)={O'}^\dagger(0,\sigma)=c_{\sigma}$, and
$O'(\sigma,\sigma ')= c^\dagger_{\sigma'}c_{\sigma}$. It is the
reaction of the conduction electron bath to these local distortions
which will renormalize the hopping amplitude as we go to lower
energies.

We can now trace out the conduction electron degrees of freedom.
The complication in evaluating each transition amplitude lies in the
fact that, it has a history dependent potential.
This can be dealt with most conveniently through the bosonization
procedure.\cite{Schotte,Solyom}
The relevant details of the bosonization applied to the impurity
problems is summarized in Appendix \ref{sec:bos}. For our impurity
problem, we need to retain only the S-wave component of the conduction
electrons. The conduction electron operator is represented by

\begin{eqnarray}
c_{\sigma} (x) = {1\over \sqrt{2\pi a}} e^{-i \Phi_{\sigma}(x)}
\label{bos.fermion}
\end{eqnarray}
where the $\Phi_{\sigma}$ field is defined in Appendix \ref{sec:bos}.
The radial dimension $x$ has been extended to the negative
half-axis. Hence, only one chiral component of the Tomonaga boson
is introduced.

The projected Hamiltonian transforms into,

\begin{eqnarray}
H_{\alpha} = H_c+ E_{\alpha}'+ \sum_{\gamma} {\delta_{\alpha}^{\gamma}
\over \pi \rho_o} (  {d\Phi_{\gamma} \over d x})_{x=0}
\label{boson}
\end{eqnarray}
where $\rho_o$ is the bare conduction electron density of states at
the Fermi level. The phase shifts $\delta_{\alpha}^{\gamma}$ are
determined by the scattering potentials defined in Eq. (\ref{potential}),

\begin{eqnarray}
&&\delta_{\sigma}^{\gamma} = \delta_1 + \delta_2
\delta_{\gamma \sigma}\nonumber\\
&&\delta_{0}^{\gamma} = 0
\label{phaseshift1}
\end{eqnarray}
where

\begin{eqnarray}
&&\delta_1=tan^{-1}(\pi \rho_o V_1/N)\nonumber\\
&&\delta_2=tan^{-1}(\pi\rho_o V_2^{\parallel}/N )
\label{phaseshift12}
\end{eqnarray}

All the short time dynamics (for $\tau$ smaller than $\xi_o$) are
included in the shift of the atomic level, leading to the renormalized
levels $E_{\alpha}' =E_{\alpha} + \Delta E_{\alpha}$. The details for
this level shift are given in Appendix \ref{sec:bos}.   The history
dependent potential is treated through introducing a canonical
transformation\cite{Schotte} at each imaginary time

\begin{eqnarray}
U_{\delta} = exp ( i \delta \Phi )
\label{can.trans2}
\end{eqnarray}
The transformed potential is time independent, due to the following
property:

\begin{eqnarray}
U_{\delta}^{\dagger}H_c U_{\delta} = H_c+{\delta \over \pi \rho_o} ({d \Phi
\over d x})_{x=0}
\label{can.trans1}
\end{eqnarray}

It is important to note that, the unitary transformation operator
given in Eq. (\ref{can.trans2}) has the same form as the fermion
operators. The transition amplitude now reduces to\cite{note3}

\begin{eqnarray}
A(\alpha;\tau_n,...,\tau_1)=&&Z_c
\sum_{\alpha_{n+1}=\alpha_1=\alpha,\alpha_2, ..., \alpha_n}
y'_{\alpha_{n+1}, \alpha_n}...
y'_{\alpha_{i+1}, \alpha_i }...
y'_{\alpha_{2}, \alpha_1 }\nonumber\\
&&\times exp[-E_{\alpha}'(\tau_1-\tau_n)-\sum_{i=2}^{n-1}
E_{\alpha_{i+1}}' (\tau_{i+1}-\tau_i)] \times \nonumber\\
&&\times <O(\alpha_{n+1}, \alpha_n ) (\tau_n)...
O(\alpha_{i+1}, \alpha_i ) (\tau_i)...
O(\alpha_2, \alpha_1 ) (\tau_1)>
\label{pert5.2}
\end{eqnarray}
Here

\begin{eqnarray}
O(\alpha_{i+1},\alpha_i) (\tau_i) \equiv
exp(H_c \tau_i) O(\alpha_{i+1},\alpha_i) exp(-H_c \tau_i)
\label{Otau}
\end{eqnarray}
where the disorder operators are given by
\begin{eqnarray}
O(\alpha_{i+1},\alpha_i) = ( \Pi_{\gamma}
U_{\delta_{\alpha_{i+1}}^{\gamma}}) O' (\alpha_{i+1},\alpha_i)
(\Pi_{\gamma} U_{\delta_{\alpha_{i}}^{\gamma}}^\dagger)
\label{O}
\end{eqnarray}
It is the product of the canonical transformation operators, which
represent the physics of a time dependent potential, and the conduction
electron operators reflecting the distortion in the conduction
electron sea. In terms of the boson fields,

\begin{eqnarray}
&&O(0,\sigma)=exp(i(1-\delta_2/\pi)\Phi_{\sigma})
exp(-i\sum_{\gamma}(\delta_1/\pi)\Phi_{\gamma})\nonumber\\
&&O(\sigma,\sigma')=exp(i(1-\delta_2/\pi)\Phi_{\sigma'})
exp(-i(1-\delta_2/\pi)\Phi_{\sigma})
\label{kinkoperator}
\end{eqnarray}

Eqs. (\ref{kinkoperator}) specify the initial values for the anomalous
dimension $e_{\alpha\beta}^{\gamma}$ of the disorder operator

\begin{eqnarray}
O(\alpha,\beta ) \equiv exp(i\sum_{\gamma}  e_{\alpha \beta}^{\gamma}
\Phi_{\gamma})
\label{dimension}
\end{eqnarray}
Specifically,

\begin{eqnarray}
&&e_{0,\sigma}^{\gamma}=- e_{0,\sigma}^{\gamma} =(1-{\delta_2 \over
\pi})\delta_{\gamma,\sigma}
-{\delta_1 \over \pi} \nonumber\\
&&e_{\sigma,\sigma'}^{\gamma}=-(1-{\delta_2 \over \pi})
\delta_{\gamma,\sigma}
+ (1-{\delta_2 \over \pi})\delta_{\gamma,\sigma '}
\label{charge}
\end{eqnarray}
\indent

Combining Eqs.(\ref{pert1}) and (\ref{pert5.2}), we arrive at

\begin{eqnarray}
{Z \over Z_o}
= \sum_{n=0}^{\infty}~ ~&& \sum_{\alpha_{n+1}=\alpha_1,...,\alpha_n}
\int_{\xi_0}^{\beta-\xi_0}{d\tau_n \over \xi_o}
...\int_{\xi_0}^{\tau_{i+1}-\xi_0} {d\tau_i \over \xi_o}
...\int_{\xi_0}^{\tau_{2}-\xi_0} {d\tau_1 \over \xi_o} ~ ~A
(\tau_n,...,\tau_1)\nonumber\\
&&y_{\alpha_{n+1}, \alpha_n}...
y_{\alpha_{i+1}, \alpha_i }...
y_{\alpha_{2}, \alpha_1 }~ ~exp[-\sum_{i}h_{\alpha_{i+1}}^o
{(\tau_{i+1}-\tau_i) \over \xi_o}]
\label{pert7}
\end{eqnarray}
where the transition amplitude is given by

\begin{eqnarray}
A (\tau_n,...,\tau_1) =&& <O(\alpha_{n+1}, \alpha_n ) (\tau_n)...
O(\alpha_{i+1}, \alpha_i ) (\tau_i)...
O(\alpha_2, \alpha_1 ) (\tau_1)> \nonumber\\
=&& {1 \over Z_c} \int exp (-S_c+ \int_0^{\beta} D \tau'
\sum_{m}j_{m}(\tau') \Phi_{\gamma} (\tau') )\nonumber\\
=&& exp (-\int D \tau' D \tau '' \sum_{\gamma} j_{\gamma} (\tau')
j_{\gamma} (\tau'') <\Phi_{\gamma} (\tau') \Phi_{\gamma} (\tau'') > )
\label{per8}
\end{eqnarray}
with the ``source current''

\begin{eqnarray}
j_{\gamma}(\tau) = &&{\sum }_{i =1}^n \delta(\tau-\tau_i)
e_{\alpha_{i+1},\alpha_i}^{\gamma}
\label{pert8.2}
\end{eqnarray}

In Eq. (\ref{pert7}), the cutoff is explicitly written in the
integration range. Also the dimensionless form of the flipping
amplitude has been introduced:
$y_{\alpha_{i+1}\alpha_{i}}=y'_{\alpha_{i+1}\alpha_{i}} \xi_o$.
Specifically, the charge and spin fugacities are given as follows,

\begin{eqnarray}
&&y(0,\sigma)=y_t=t\xi_o\nonumber\\
&&y(\sigma \ne \sigma ') = y_j = {V_2^{\perp} \over N } \xi_o
\label{fugacity}
\end{eqnarray}

In deriving Eq. (\ref{pert7}), we have also absorbed in $Z_0$, besides
the free conduction electron partition function $Z_c$, a shift in the
ground state energy such that $\sum_{\alpha}h_{\alpha}^o =0$.
Specifically,

\begin{eqnarray}
h_0^o &&=-E_d' \xi_0 {N \over (N+1)}\nonumber\\
h_{\sigma}^o &&=E_d' \xi_0 {1\over (N+1)}
\label{fields}
\end{eqnarray}

Using the ${1 \over \tau}$ behavior\cite{finiteT} of the long time
correlation function for the Tomonaga bosons, we finally arrive at

\begin{eqnarray}
{Z \over Z_0}
= \sum_{n=0}^{\infty} \sum_{\alpha_{n+1}=\alpha_1,...,\alpha_n}
\int_{\xi_0}^{\beta-\xi_0}{d\tau_n \over \xi_o}
...\int_{\xi_0}^{\tau_{i+1}-\xi_0} {d\tau_i \over \xi_o}
...\int_{\xi_0}^{\tau_{2}-\xi_0} {d\tau_1 \over \xi_o}
exp(-S[\tau_1, ... \tau_n ] )
\label{sumoverhis}
\end{eqnarray}
where

\begin{eqnarray}
S[\tau_1, ... \tau_n ] = &&\sum_{i<j}\sum_{\sigma}
(e_{\alpha_i\alpha_{i+1}}^{\sigma})
(e_{\alpha_j\alpha_{j+1}}^{\sigma})
ln {(\tau_j - \tau_i) \over \xi_o}\nonumber\\
&&- \sum_i ln (y_{\alpha_i\alpha_{i+1}})
+\sum_{i}h_{\alpha_{i+1}} {(\tau_{i+1}-\tau_i) \over \xi_o}
\label{hisaction}
\end{eqnarray}

The partition function is now a summation over all possible histories
of the local degrees of freedom which fluctuate between the $N+1$
local states $|\alpha >$. Each history, labeled by
$\{\alpha_1,...,\alpha_n; \tau_1,...,\tau_n\}$, is a sequence of
transitions between the local states from $\alpha_1$ through $\alpha_n$
taking place at the time $\tau_1<...<\tau_n$,
as is illustrated in Fig. \ref{hop}. The action Eq.
(\ref{hisaction}) gives the statistical weight of such a history.

We can interpret Eq. (\ref{hisaction}) as that associated with the
partition function of a plasma of kinks with logarithmic interactions.
It has (multi-component) charge $e_{\alpha\beta}^\sigma$ and fugacity
$y_{\alpha \beta}$.

\subsection{Equivalence to a Discrete Spin Chain with Long Range
Interactions}
\label{sec:parfun2}

The partition function now has its action expressed as a summation
over logarithmic interactions between the pair-wise kinks. Each kink
is a flip in the original `spin' space (i.e., the local states
$|\alpha>$). The logarithmic interactions between the kink-pairs can
therefore be transformed into interactions between the ``spins''.
It is straightforward to show that, the action in
Eq. (\ref{hisaction}) can be written in a symmetric form,

\begin{eqnarray}
S[\tau_1, ... \tau_n ] = &&\sum_{i<j}
(K(\alpha_i, \alpha_j) + K(\alpha_{i+1}, \alpha_{j+1})
- K(\alpha_i, \alpha_{j+1}) - K(\alpha_{i+1}, \alpha_{j}) )
ln {(\tau_j - \tau_i) \over \xi_o}\nonumber\\
&&- \sum_i ln (y_{\alpha_i\alpha_{i+1}})
+\sum_{i}h_{\alpha_{i+1}} {(\tau_{i+1}-\tau_i) \over \xi_o}
\label{hisaction2}
\end{eqnarray}
which can then be transformed into a long range interaction between the
spins,

\begin{eqnarray}
S[\tau_1, ... \tau_n ] = &&\sum_{i<j} K(\alpha_i,
\alpha_j) {\xi_o^2 \over (\tau_j-\tau_i)^{2}}\nonumber\\
&&- \sum_i ln (y_{\alpha_i\alpha_{i+1}})
+\sum_i h_{\alpha_{i+1}} {(\tau_{i+1}-\tau_i) \over \xi_o}
\label{spinaction}
\end{eqnarray}
where the spins $|\alpha>$ can be in
$N+1$ components.

This transformed action is a special case of a class of
discrete spin models with $1/r^2$ interactions considered by
Cardy,\cite{Cardy} generalized to incorporate an effective field.
The bare values of the stiffness constants $-K(\alpha,\beta)$ are
determined by the bare values of the charges
$e_{\alpha\beta}^\sigma$,

\begin{eqnarray}
K(\alpha, \beta) = -{1 \over 2} \sum_{\gamma}
(e_{\alpha,\beta}^{\gamma})^2
\label{stiffness}
\end{eqnarray}
Specifically,

\begin{eqnarray}
&&K(0,\sigma) \equiv -\epsilon_t \nonumber\\
&&K(\sigma,\sigma') \equiv -\epsilon_j(1-\delta_{\sigma,\sigma'})
\label{stiffness2}
\end{eqnarray}
where the bare values of the charge and spin stiffness constants are

\begin{eqnarray}
&&\epsilon_t^o={1 \over 2}[ ( 1 - {\delta_2 \over {\pi}} - {\delta_1
\over {\pi}})^2+ ( N -1 ) ( {\delta _1 \over {\pi}})^2 ]\nonumber\\
&&\epsilon_j^o= ( 1 - {\delta_2 \over {\pi} } )^2
\label{stiffness3}
\end{eqnarray}

\bigskip
\section{RG Flow of the Generalized Anderson Model\\
and the Nature of the Fixed Points}
\label{sec:flow}

In this section, we derive the RG equations using the atomic
representation of the partition function derived in the previous
section. The RG flow has the most general form in terms of the charge
and spin fugacities, the charge and spin stiffness constants, and the
$d$-level (a symmetry breaking field). The flow equations enable us to
identify all the fixed points. They correspond to a universal strong
coupling fixed point, a plane of weak coupling fixed points, and a
plane of intermediate coupling fixed points. All these fixed points
are studied in detail. We will establish in the next section that
these fixed points describe Fermi liquid and non-Fermi liquid phases
of the extended Hubbard model.

\subsection{Renormalization Group Equations}
\label{sec:flow1}

We construct RG equations for small $E_d$ in this section. As was
discussed in Section \ref{sec:ehm3}, in certain range of parameters,
the extended Hubbard model in infinite dimensions maps onto an
impurity model which has a zero renormalized mass ($d$ level $E_d$).
Therefore, the small $E_d$ assumption is justified. However,  since
the RG procedure we use is perturbative in fugacities but
non-perturbative in stiffness constants and fields, the RG equations
for finite fields can also be constructed and are given in the
Appendix \ref{sec:scaling.h}.

The RG equations describe the flow of the dimensionless couplings as
the bandwidth $1 /\xi$ is reduced. The details of the derivations
are presented in Appendix \ref{sec:scaling.h}, where arbitrary local
states with a partition function of the form given in Eq.
(\ref{sumoverhis}-\ref{spinaction}) are considered.
For our $N+1$ local state problem, the RG charges are the charge and
spin fugacities $y_t$ and $y_j$ (defined in Eq. (\ref{fugacity})), the
charge and spin stiffness constants $\epsilon_t$ and $\epsilon_j$ (defined in
Eqs. (\ref{stiffness2}) and (\ref{stiffness3})), and the dimensionless
$d$ level $E_d\xi=h_{\sigma}-h_0$. The RG equations are given as follows,

\begin{eqnarray}
&&{d y_t \over {d ln \xi}} =
(1- \epsilon_t)y_t + ( N -1 ) y_t y_j\nonumber\\
&&{d y_j \over {d ln \xi} }=
(1- \epsilon_j)y_j + ( N -2 ) y_j^2 + y_t^2\nonumber\\
&&{ d \epsilon_t  \over {d ln \xi}}=
-2\epsilon_t (N+1) y_t^2
+\epsilon_j (N-1) (y_t^2 - y_j^2)\nonumber\\
&&{ d \epsilon_j \over {d ln \xi }}
= -2\epsilon_j (y_t^2+Ny_j^2)\nonumber\\
&&{ d {E_d\xi} \over {d ln \xi}}= (N-1) (y_t^2 - y_j^2 )
+E_d \xi (1-(N+1)y_t^2)\nonumber\\
&&{ d {F\xi} \over {d ln \xi}}= F \xi
-{2N \over N+1} y_t^2-{N(N-1) \over N+1} y_j^2
\label{scaling}
\end{eqnarray}
where $F$ is the free energy.\cite{rotat-inv}

In the renormalization of the fugacities, the linear terms give the
associated anomalous dimensions, while the quadratic terms reflect the
non-abelian nature of our $N+1$ state problem. Physically, these cross
terms reflect the coupling between the local spin and charge degrees of
freedom. The renormalization of the stiffness constants reflect the
correction to the interactions due to the fugacities of the charge and
spin kinks.

In the renormalization of the energy level, the $y_t^2$ term arises
because in integrating out the degrees of freedom the empty state
gains a hybridization energy $N$ times as much as a spin state does.
This reflects the particle-hole asymmetry of the mixed valence
problem.\cite{Haldane} On the other hand, a spin state gains
exchange energy through the other $(N-1)$ spin states, leading to the
$y_j^2$ term.
We emphasize that, the symmetry breaking field is a relevant
perturbation.

For N=1, the above RG equations reduce to those of the anisotropic
Kondo problem, through identifying $t$ with ${1 \over 2} J_{\perp}$,
$\epsilon_t$ with $(1-2/\pi tan^{-1}(J_{\parallel}\xi_o/4))^2$, and
$-E_d$ with the magnetic field $H$ applied on the local
moment.\cite{Wieg,Schlot} For N=2 and in the limit $V_1=V_2=0$, we
reproduce equations of Haldane's \cite{Haldane} for the flow of the
$d$ level and the hybridization for the asymmetric Anderson model in
the mixed valence regime.

\subsection{Fixed Points and Critical Behavior}
\label{sec:flow2}

The kinks are bound together when the amount of attraction between
them are strong enough. This leads to fugacities renormalized to zero.
As in the Kosterlitz-Thouless transition in the XY
magnet,\cite{Kosterlitz} unbinding of the kinks occur when the
attractions between the kinks become weak. In our case, $\epsilon_t$
and $\epsilon_j$ reflect the strength of the long range interactions
between the kinks. Since the spin and charge kinks are coupled, the
unbinding of these kinks can occur either at the same time or at different
stages. The above RG equations lead to three kinds of fixed points.
The phase diagram in the $\epsilon_t-\epsilon_j$ plane is given in
Fig. \ref{pdeps}.

The weak coupling fixed points occur when both the spin and charge kinks
are bound, corresponding to the hybridization and the exchange
coupling both being irrelevant. This occurs when both the renormalized
charge and spin stiffness constants are large enough,

\begin{eqnarray}
&&\epsilon_t^*>1\nonumber\\
&&\epsilon_j^*>1
\label{weak}
\end{eqnarray}
where the stiffness constants are renormalized from the bare values by
the fugacities. To second order in the fugacities,

\begin{eqnarray}
&&\epsilon_t^*=\epsilon_t^o-(y_t^o)^2{{2(N+1)\epsilon_t^o
-(N-1)\epsilon_j^o} \over {2(\epsilon_t^o-1)}}
-(y_j^o)^2 {{(N-1) \epsilon_j^o} \over {2(\epsilon_j^o-1)}}\nonumber\\
&&\epsilon_j^*=\epsilon_j^o-(y_t^o)^2{ \epsilon_j^o \over
\epsilon_t^o-1}
- N (y_j^o)^2{\epsilon_j^o \over {\epsilon_j^o-1}}
\label{epsilon.ren}
\end{eqnarray}
For N=1, this reduces to $\delta_1/\pi <
-(\sqrt{2}-1)-t/\Gamma$.\cite{Wieg,Schlot}
For $N \ge 2$, it corresponds to

\begin{eqnarray}
&&N ({\delta_1 \over {\pi} })^2- 2 ( 1-
{\delta_2 \over{\pi}} ) {\delta_1 \over{\pi}}
- (2- ( 1- {\delta_2 \over{\pi}} )^2) > 0\nonumber\\
&&{\delta_2 \over {\pi}} < 0
\label{irrelevant}
\end{eqnarray}

The strong coupling mixed valence fixed point occurs when both
the spin and charge kinks are unbound, corresponding to the
hybridization and the exchange coupling both being relevant.
The basin of attraction is given by

\begin{eqnarray}
\epsilon_t^* < 1
\label{strong}
\end{eqnarray}
and a range of $\epsilon_j^* < 1$ when $\epsilon_t^* > 1$.

This strong coupling fixed point is beyond the reach of perturbative
RG. It is however in the same universality class as the usual strong
coupling phase of the Anderson model: below a non-zero renormalized
Fermi energy, both the local spin and charge degrees of freedom are
quenched through the formation of quasiparticles.

In the spinful problem ($N \ge 2$), for a range of couplings within
the following domain

\begin{eqnarray}
&&\epsilon_t^*>1\nonumber\\
&&\epsilon_j^*<1
\label{inter}
\end{eqnarray}
an intermediate phase can occur for which the spin kinks are unbound
while the charge kinks are bound. Here the hybridization is
irrelevant, while the exchange coupling is relevant. The
local spins are quenched below a non-zero coherence energy,
while the local charges are not. We note that, the RG equations imply
that, when $y_j$ becomes very
large, it will start to drive $y_t$ to increase. However,
this late stage is beyond the reach of perturbative RG. We will
establish in Section \ref{sec:flow5} that, the fixed point with
irrelevant $y_t$ and relevant $y_j$ is indeed stable. The transition
to the strong coupling phase occurs when the effective charge
stiffness constants, further renormalized due to the unbound spin-kink
plasma,  reaches the critical value 1.

We emphasize that, from the quadratic couplings in the scaling
equations, a relevant charge coupling necessarily drives the spin
coupling to be relevant, even with a bare $\epsilon_j>1$. Physically,
a spin kink (created by $d_{\sigma}^\dagger d_{\bar{\sigma}}$) is the
composite of two charge kinks (created by $d_{\sigma}$). Unbounded
charge kinks will strongly screen the interactions between the spin
kinks, leading to the unbinding of the latter as well.

The phase boundaries are given in Fig. \ref{pdeps}. The vertical thick
line separates the weak coupling phase from the strong coupling phase.
The horizontal thick line separates the weak coupling phase from the
intermediate coupling phase. The (schematic) dashed line represents
the boundary between the intermediate coupling and strong coupling
phases.

We note the analogy of these three phases with those in the two
dimensional defect-driven melting problem.\cite{Nelson} There, the
binding-unbinding transitions are associated with the dislocation and
the disclination defects. A dislocation is the composite of a pair of
disclinations, in analogy to the spin kink being a composite of a pair
of charge kinks. The hexatic phase exhibits bound disclinations and
unbound dislocations, in analogy to our intermediate coupling phase
with irrelevant charge fugacity and relevant spin fugacity.

In terms of the physical interactions, i.e. the density-density
interaction $V_1$ and the exchange interaction $V_2$,
the basin of attraction of the weak coupling fixed points corresponds
to a range of attractive density-density and ferromagnetic exchange
interactions, while the intermediate coupled fixed points occur over a
range of attractive density density and antiferromagnetic Kondo
exchange interactions. We stress that, these are effective
interactions at the beginning of the scaling trajectory. In
generic models in infinite dimensions with a non-Lorentzian density
of state, there are high energy dynamics which have to be integrated
out before the low energy scaling regime is reached. We will discuss
the implications of our results for more realistic models in
Section \ref{sec:conclu}.

We end this subsection with a discussion of the critical behavior of
the zero temperature quantum phase transitions between these various
states we identified.
For reasons to be described at the end, these zero temperature
quantum phase transitions can be formulated only in infinite
dimensions. At finite dimensions we expect that the system will have
sharp crossovers.

The transitions are characterized by the collapse of an
energy scale,

\begin{eqnarray}
\epsilon_F \sim \epsilon_0 {\rm e}^{-{1 \over (\epsilon -
\epsilon_c)^{\eta}}}
\label{critical}
\end{eqnarray}

The transition between the weak coupling and strong coupling
mixed valence states takes place when we vary the interactions (and
hence $\epsilon_t^*$ and $\epsilon_j^*$) through the vertical thick line
in Fig. \ref{pdeps}. This one stage transition is illustrated in Fig.
\ref{criti}(a), where $\epsilon_1$ labels a line in the
$\epsilon_t-\epsilon_j$ plane that passes through the vertical thick
line at $\epsilon_c$. The RG equations imply that, the unbinding
transition for spin kinks is driven by that of the charge kinks. The
critical behavior is therefore determined by the renormalization of
$y_t$ only. Hence, $\epsilon_F$ in Eq. (\ref{critical}) corresponds to
the Fermi energy of the Fermi liquid in the strong coupling state, and
the exponent $\eta={1 \over 2}$.

When we vary the interactions (and hence $\epsilon_t$ and
$\epsilon_j$) through the horizontal thick line and the dashed line in
Fig. \ref{pdeps}, two stages of transition take place, from the weak
coupling through the intermediate coupling to the strong coupling
phase. This is illustrated in Fig. \ref{criti}(b), where $\epsilon_2$
labels a line in the $\epsilon_t-\epsilon_j$ plane that passes through
the horizontal thick line at $\epsilon_{c2}$, and through the dashed
line at $\epsilon_{c1}$. Near $\epsilon_{c2}$, the $\epsilon_F$ in Eq.
(\ref{critical}) refers to the coherence energy for coherent spin
excitations in the intermediate coupling state. Close to
$\epsilon_{c1}$, on the other hand, $\epsilon_F$ refers to the
coherence energy for the coherent charge excitations
in the strong coupling phase. The exponent
characterizing the unbinding of spin kinks at $\epsilon_{c2}$ is given
by $\eta=1$. Due to rotational invariance, it differs from the
exponents in the corresponding $N$-component Coulomb gas problem, and
is independent of N.
The exponent characterizing the unbinding of charge kinks at
$\epsilon_{c1}$, on the other hand, is $\eta={1 \over 2}$.

In the following subsections, we discuss in detail the nature of these
individual phases.

\subsection{Strong Coupling Mixed Valence State}
\label{sec:flow4}

Within the strong coupling mixed valence state, both the hybridization
and the exchange coupling are relevant. The effect of interactions
becomes smaller as we go to lower energies. The nature of
the fixed point is that of the strong coupling mixed valence state of
the usual Anderson model. The ground state is well described by
various variational wavefunctions, such as those of Varma and
Yafet.\cite{MV} The physics of this state can be systematically
studied in the slave boson condensed phase in the large N
approach.\cite{slaveboson} Within the strong coupling phase,
there exist the empty orbital, local moment and mixed valence
regimes.\cite{Haldane,Krishna} Crossover between these various regimes
occur as the renormalized $d$-level is varied with respect to the
Fermi level.

In the following, we use our scaling equations to determine a
generalized form for this $d$-level shift and resonance width
$\Delta(\xi)=\pi y_t^2(\xi)/\xi$, in the presence of the finite
density-density and exchange interactions.
In the generic case, the flow of the $d$-level can only be derived
through integrating numerically the closed set of the RG flow.  For a
vanishing bare spin fugacity, $y_j^0=0$, and when the renormalization
of the stiffness constants is ignored, the leading terms for the
running $d$-lvel and the running resonance width can be determined from

\begin{eqnarray}
{d E_d \over {d ln \xi}}~&&=~ (N-1) {\Delta_o \over \pi} ({\xi \over
\xi_o})^{-(2\epsilon_t-1)} \nonumber\\
{d \Delta \over {d ln \xi}}~&&=~ (1-2\epsilon_t) \Delta_o
({\xi \over \xi_o})^{-(2\epsilon_t-1)}
\label{scaling.s1}
\end{eqnarray}
where $\Delta_o= \pi (y_t^o)^2/\xi_o$. This leads to

\begin{eqnarray}
E_d (\xi ) &&= E_d' + (N-1){\Delta_o \over \pi}
{{1-(\xi / \xi_o)^{-(2 \epsilon_t-1)}}\over  {2 \epsilon_t-1}}\nonumber\\
\Delta(\xi) &&= \Delta_0 (\xi / \xi_0 )^{-(2 \epsilon_t - 1)}
\label{scaling.s2}
\end{eqnarray}
 from which we construct the invariant level and resonance width

\begin{eqnarray}
E_d^* &&= E_d' + (N-1) {\Delta_o \over \pi}
{1-(\Delta_0 \xi_o)^{-{2 \epsilon_t -1 \over 2(1-\epsilon_t )}} \over
2 \epsilon_t-1}\nonumber\\
\Delta^* &&= \Delta_0 (\Delta_0 \xi_0 )^{{2 \epsilon_t-1 \over
2(1-\epsilon_t )}}
\label{scaling.s3}
\end{eqnarray}
In terms of these scaling invariants, the running $d$ level and
resonance width are given by

\begin{eqnarray}
E_d (\xi ) &&= E_d^* + (N-1) {\Delta^* \over \pi}
{1- (\xi \Delta^*)^{-(2\epsilon_t-1)} \over  2\epsilon_t-1}\nonumber\\
\Delta(\xi) &&= \Delta^* (\xi \Delta^* )^{-(2\epsilon_t-1)}
\label{scaling.s4}
\end{eqnarray}
A logarithmic shift is recovered when $\epsilon_t=\epsilon_t^0=1/2$ is
used, corresponding to $V_1 = V_2 =0$.

At low temperatures, the crossover between the various regimes can be
determined from comparing the invariant $d$ level $E_d^*$ and the
invariant resonance width $\Delta^*$. The local moment and empty
orbital regimes occur when $E_d^*<<-\Delta^*$ and  $E_d^*>>\Delta^*$
respectively.  For $|E_d^*|<\Delta^*$, the system is in the mixed
valence regime. This crossover behavior is illustrated in Fig.
\ref{crossover}(a). Here we have used Eq. (\ref{dlevel}), so that the
variation of $E_d^*$ is described in terms of the chemical potential
$\mu$ of the lattice model. At zero temperature, the mixed valence
crossover extends over a scale of $\sim \Delta^*$. This crossover can
also be illustrated through the $n$ vs $\mu$ curve, shown
schematically in Fig. \ref{nmu}(a) and to be discussed further in
Section \ref{sec:phases}.

Finite temperature (or frequency) cuts off the scaling. As
temperature is increased, both the running $d$-level $E_d (T)$ and the
running resonance width $\Delta(T)$ vary. The crossover temperature
scale, separating the high temperature mixed valence regime from the
low temperature empty orbital or local moment regime, can be
determined from $E_d(T)={\rm max} ( T, \Delta(T))$. This specifies the
crossover lines$-$the dashed lines$-$in Fig.
\ref{crossover}(a).\cite{Haldane} In the local moment regime, there
will be a further crossover at the scale of the Kondo temperature (not
shown in the figure). In all these regimes, the low energy behavior is
described by the Fermi liquid theory. The renormalized Fermi energy is
given by the renormalized resonance width $\Delta^*$.

\subsection{Weak Coupling Mixed Valence State}
\label{sec:flow3}

At the weak coupling fixed points, the fugacities become smaller as we
go down to lower energies. Therefore, the atomic expansion in terms of
the running fugacities become more justified at lower energies. All
the correlation functions can be derived from the RG procedure.

The running fugacities and $d$ level can be derived from integrating
the RG flow equations:

\begin{eqnarray}
&&t(\omega) = t (\omega \xi_o )^{\epsilon_t^*}\nonumber\\
&&V_2^{\perp} (\omega )=V_2^{\perp} (\omega \xi_o)^{\epsilon_j^*}\nonumber\\
&&E_d ( \omega )= E_d^{\ast} - {A \over \xi_o} (\omega \xi_o
)^{2\epsilon_t^* -1}
\label{w1}
\end{eqnarray}
where $A={(N-1){y_t^o}^2 \over {2\epsilon_t^*-1}}$, and
$E_d^{\ast}= E_d ' + {A \over \xi_o}$.
In this case, the $d$-level shift due to the particle-hole asymmetry
is small, since the hybridization is irrelevant.

The crossover diagram is given in Fig. \ref{crossover}(b). There again
exist  local moment, empty orbital, and mixed valence regimes. The
important difference with the the crossover  diagram
of the strong coupling phase (Fig. \ref{crossover}(a))
is that, in this
case the renormalized resonance width $\Delta^*$ vanishes at zero
temperature; A zero temperature critical point occurs at $\mu=\mu_c$.
This can also be shown through the $n$ vs $\mu$ curve given in Fig.
\ref{nmu}(b), where $n$ is the total electron density.

To determine the finite temperature crossovers between the different
regimes, we define the running
``resonance width'' $\Delta (T)$ defined by

\begin{eqnarray}
\Delta ( T) = \pi y_t^2 T = \pi t^2\xi_o (T \xi_o
)^{(2\epsilon_t^*-1)}
\label{w2}
\end{eqnarray}
The local moment, empty orbital, and the mixed valence regimes occur
when $E_d(T) << - {\rm max} (T,\Delta (T))$,
$E_d(T) >> {\rm max} (T, \Delta ( T))$, and $ |E_d(\xi)|<
{\rm max} (\Delta (T )$ respectively. This determines the crossover
behavior illustrated in Figs. \ref{crossover}(b) and \ref{nmu}(b).

For $ \sim (-A/\xi_o) <E_d^*< \sim A/\xi_o$, an additional crossover
occurs at an energy scale $T'$,

\begin{eqnarray}
T'= {1\over \xi_o} ({|E_d^*|\xi_o \over A})^{1 \over {2\epsilon_t^*-1}}
\label{w3}
\end{eqnarray}
which is determined from $| E_d (T ') | \sim \Delta ( T' )$.
The system is in the critical regime at temperatures higher than $T'$,
since here it is essentially an $E_d (T)=0$ problem, with a
temperatur-dependent resonance width which is smaller than the
temperature. In particular, at $E_d^*=0$, i.e. $\mu=\mu_c$,  this
mixed valence critical regime persists all the way down to zero
temperature. The mixed valence state corresponds to a massless
critical point.

The mixed valence phase here is characteristically different from the
strong coupling mixed valence phase. The irrelevant hybridization
makes the running resonance width to be always smaller than the scale,
$\Delta(\omega) <\omega$. Local correlation functions in the mixed
valence regime can be calculated through the RG
procedure.\cite{Kosterlitz}

Here we focus on the local Green's functions, which is a $2 \times 2$
matrix for each spin component:

\begin{eqnarray}
G = \left( \begin{array}{ll}
                         G_{cc}~ ~ ~ ~ ~ ~ ~ ~ ~ &
G_{cd}\\
                         G_{dc}~ ~ ~ ~ ~ ~ ~ ~ ~ &
G_{dd}
                      \end{array}
\right)
\label{greenfun}
\end{eqnarray}

The details of the RG calculation for the
renormalization of the exponents can be found in Appendix
\ref{sec:corr}. The results are as follows. $G_{dc}$ and $G_{dd}$
exhibit infrared divergences, and have the following low energy
behavior,

\begin{eqnarray}
&&G_{dc} (\omega) \sim \omega^{-1+\alpha}\nonumber\\
&&G_{dd} (\omega) \sim \omega^{-1+\beta}
\label{corr10}
\end{eqnarray}
Here the exponents are, to leading order, as follows

\begin{eqnarray}
&&\alpha=-(\delta_1^*+\delta_2^*)/\pi+((\delta_1^*+\delta_2^*)/\pi)^2
+(N-1)(\delta_1^*/\pi)^2\nonumber\\
&&\beta=((\delta_1^*+\delta_2^*)/\pi)^2 +(N-1)(\delta_1^*/\pi)^2
\label{corr11}
\end{eqnarray}
where $\delta_{1,2}^*$ are the renormalized phase shifts determined
 from Eq. (\ref{epsilon.ren}). The algebraic divergences in $G_{dc}$
and $G_{dd}$ reflect the fact that, both the impurity spin and charge
degrees of the freedom are not quenched.

On the other hand, the local $c$-electron Green's function $G_{cc}$
does not show infrared divergences. Formally, creating a local
$c$-electron does not creat any kink in the path integral
representation. Hence, $G_{cc}$ is not renormalized from its fixed point
form. At the mixed valence fixed point, $G_{cc}$
has a long time
${1\over \tau}$ behavior.  The absence of infrared divergences in the
electron bath spectral function will be used in Sec. \ref{sec:phases4}
to argue that, our results derived using a Lorentzian density of
states can be applied to the case of generic density of states as well.

\subsection{Intermediate Coupling Mixed Valence State}
\label{sec:flow5}

In this case, the RG equations yield an initially decreasing charge
fugacity $y_t$ and an initially increasing spin fugacity $y_j$. When
$y_j$ grows to order unity, the flow equations would indicate that
$y_t$ starts to grow. However, this is beyond the regime of validity
for the perturbative RG.

To understand qualitatively the nature of the fixed point, we analyze
the model in the limit that the antiferromagnetic exchange $V_2$
is much larger than the conduction electron band width, while the
hybridization $t$ is much smaller than the band width. This serves as
an effective Hamiltonian that the original Hamiltonian renormalizes to
at an intermediate scale.

\begin{eqnarray}
\tilde{H}_{eff} =&& V_2 \sum_{\sigma,\sigma'} d_{\sigma}^\dagger d_{\sigma'}
c_{o,\sigma'}^\dagger c_{o,\sigma}
+V_1 \sum_{\sigma,\sigma'} d_{\sigma}^\dagger d_{\sigma}
c_{o,\sigma'}^\dagger c_{o,\sigma'}
+ \epsilon_d' \sum_{\sigma} d_{\sigma}^\dagger d_{\sigma}
+ {U \over 2} \sum_i \sum_{\sigma \ne \sigma'}^N d^{\dagger}_{i\sigma}
d_{i\sigma } d^{\dagger}_{i\sigma ' } d_{i\sigma ' }\nonumber\\
&& +\sum_{\sigma} t (d_{\sigma}^\dagger c_{o,\sigma} + h.c.)
+\sum_{i_1,\sigma} t_1 (c_{1,\sigma}^\dagger c_{o,\sigma} + h.c.)
+\sum_{<i,j \ne 0>} t_{ij} (c_{i,\sigma}^\dagger c_{j,\sigma} + h.c.)
\label{i1}
\end{eqnarray}
Here $c_{1,\sigma}^\dagger$
creats effective neighboring states which are coupled to the
conduction electron state at the impurity site that $c_{0,\sigma}^\dagger$
creats. We consider N=2 for simplicity.  The parameters are such that
$V_2>>V_1,t_1 >>t$, and $U=\infty$.

To order $O({t_1 \over V_2},{t \over V_2})$, diagonalizing the $V_2$
coupling first leads to singlet and triplet subsectors within the
$n_d=1$ sector. And the triplet sector locates at high energies. The
mixed valence condition is satisfied at an $\epsilon_d'$ such that the
singlet sector and the $n_d=0$ sector are (nearly) degenerate.
We can now integrate out the high energy states. This leads to the
following effective Hamiltonian for the low energy sectors,

\begin{eqnarray}
H_{eff} =&& \epsilon_s (s^\dagger s -e^\dagger e) + \sum_k \tilde{\epsilon}_k
\tilde{c}_{k,\sigma}^\dagger \tilde{c}_{k,\sigma} \nonumber\\
&&+ \sum_{k,k',\sigma} ( {t'}_{k,k'} s^\dagger e
\tilde{c}_{k,\sigma}\tilde{c}_{k,-\sigma}sgn ( \sigma )  + h.c. )
+\sum_{k,k',\sigma} {V'}_{k,k'} (s^\dagger s - e^\dagger e )
\tilde{c}_{k,\sigma}^\dagger\tilde{c}_{k,\sigma}
\label{i2}
\end{eqnarray}
Here $s^\dagger$ and $e^\dagger$ creat the local singlet and the local
empty state respectively, while $\tilde{c}_{k,\sigma}^\dagger$ creats
conduction electron scattering states. The effective interaction and
the effective ``hybridization'' amplitude are given as follows

\begin{eqnarray}
{V'}_{k,k'} &&\sim {2 V_1 \over (V_2 - V_1)(V_2 + V_1)} t_1^2
\label{i3}\nonumber\\
{t'}_{k,k'} &&\sim {1 \over 2} t
\end{eqnarray}

Given that $t$ is small, and  $V_s' < 0$, the effective
``hybridization'' is {\it marginally irrelevant}. Therefore, the spin
and charge sectors are indeed decoupled asymptotically.

The crossover diagram is also given qualitatively by Fig.
\ref{crossover}(b), except for the local moment regime in which the
spin degrees of freedom is eventually quenched through the Kondo
process. The $n$ vs. $\mu$ curve at zero temperature is again
qualitatively given by Fig. \ref{nmu}(b).

For the mixed valence regime, the spectral functions have the form of
a convolution between the resonance in the spin sector and the
algebraic term in the charge sector, and therefore also exhibit an
algebraic behavior at low energies. The exponents can be determined
 from the effective charge stiffness constants, renormalized (further
 from $\epsilon_t^*$) due to the screening of the free spin kink
plasma. This further renormalization can be analyzed in detail by
following the scaling trajectory and using Eq. (\ref{i2}), in the same
spirit as the analysis of the hexatic phase using the Debye-H${\rm
\ddot{u}}$ckel approximation.\cite{Nelson} In particular, the exponent
is universal when
the critical line separating the intermediate coupling and the strong
coupling states, i.e. the dashed line in Fig. \ref{pdeps}, is
approached.

\bigskip
\section{\bf Non-Fermi Liquid and Fermi Liquid Phases\\
of the Extended Hubbard Model}
\label{sec:phases}

In this section, we establish the implications of the results of the
associated impurity problem in the previous two sections on the
phases in the extended Hubbard model. We show that, the phases
corresponding to the weak coupling and intermediate coupling mixed
valence states are metallic non-Fermi liquid phases. They have
vanishing quasiparticle residue, algebraic local correlation
functions, and asymptotically separated charge and spin excitations.
The phase corresponding to the strong coupling mixed valence state is
a usual Fermi liquid.

\subsection{Metallic non-Fermi Liquid: Weak Coupling\\
and Intermediate Coupling Phases}
\label{sec:phases1}

Consider first the phase corresponding to the weak coupling mixed
valence state. We first establish that, the critical chemical
potential $\mu_c$, at which the mixed valence state persists to zero
temperature, corresponds to {\it a range of electron densities.}

The local correlation functions of the extended Hubbard
model in infinite dimensions are given by the impurity problem. In
particular, the occupation numbers in the lattice, for a given
chemical potential, can be obtained from the local Green's functions
of the corresponding impurity model. Therefore, our analyses in
Sections \ref{sec:ehm3} and \ref{sec:flow3} establish that, at zero
temperature $n_d$ and also $n=n_d+n_c$ are discontinuous functions of
the chemical potential: for $\mu > \mu_c$, $n_d \approx 1 -
O(t^2)$ while for $\mu < \mu_c$, $n_d \approx O(t^2)$.  At finite
temperatures, the discontinuity is smoothed out and changes into a
fast crossover. Here we take $\mu =\mu_c$ which again corresponds to a
{\it finite range} of densities in the lattice model.

It is remarkable that the fact that our impurity model is associated
to a lattice problem forces the impurity model to be exactly at
criticality, with a larger symmetry than we would have naively
expected. The essential point is that, in order for an {\it incoherent
state} to be metallic, it is necessary to allow for charge transfer
between the localized degrees of freedom and the bath. This can only
happen if the local charge degree of freedom is in equilibrium with
the conduction electron bath. This requires the heavy level to be at
the chemical potential.

To determine the excitation spectrum of the system, we consider the
lattice Green's function. In the limit of infinite dimensions, all the
lattice correlation functions can be reconstructed from the local
correlation functions calculated in the associated impurity problems.
The lattice Green's function,

\begin{eqnarray}
G ({\bf k}, \tau) ~=~&&\left( \begin{array}{ll}
        G_{cc}({\bf k},\tau)~ ~ ~ ~ ~ ~ ~ ~ ~ & G_{cd}({\bf k},\tau)\\
        G_{dc}({\bf k},\tau)~ ~ ~ ~ ~ ~ ~ ~ ~ & G_{dd}({\bf k},\tau)
                      \end{array}
\right)\nonumber\\
&&=~ \left( \begin{array}{ll}
                         -<{\rm T} c_{\bf k}({\tau}) c_{\bf k}^\dagger
(0)>~ ~ ~ ~ ~~ ~ ~ ~ &
-<{\rm T} c_{\bf k} ({\tau}) d_{\bf k}^\dagger (0)>\\
                         -<{\rm T} d_{\bf k} ({\tau}) c_{\bf
k}^\dagger (0)>~ ~ ~ ~ ~ ~ ~ ~ ~ &
-<{\rm T} d_{\bf k} ({\tau}) d_{\bf k}^\dagger (0)>
                      \end{array}
\right)
\label{mnfl3}
\end{eqnarray}
can be determined from the Dyson equation for the lattice model

\begin{eqnarray}
G ( {\bf k}, i\omega_n)
&&=~\left( \begin{array}{ll}
        i\omega_n+\mu-\epsilon_{\bf k}-\Sigma_{cc}(i\omega_n) ~ ~ ~ ~
 ~ ~ ~ ~ ~ & t-\Sigma_{cd}(i\omega_n)\\
        t-\Sigma_{dc}(i\omega_n)~ ~ ~ ~ ~ ~ ~ ~ ~ & i\omega_n+\mu-\epsilon_d^o
-\Sigma_{dd}(i\omega_n)
                      \end{array}
\right)^{-1}
\label{mnfl5}
\end{eqnarray}
where the momentum-independent self-energy can be calculated
directly from the local Green's functions of the impurity model

\begin{eqnarray}
\Sigma (  i\omega_n)~&&=~\left( \begin{array}{ll}
        \Sigma_{cc}(i\omega_n)~ ~ ~ ~ ~ ~ ~ ~ ~ & \Sigma_{cd}(i\omega_n)\\
        \Sigma_{dc}(i\omega_n)~ ~ ~ ~ ~ ~ ~ ~ ~ & \Sigma_{dd}(i\omega_n)
                      \end{array}
\right)\nonumber\\
&&=~\left( \begin{array}{ll}
        (G_o^{-1})_{cc}(i\omega_n)~ ~ ~ ~ ~ & (G_o^{-1})_{cd}(i\omega_n)\\
        (G_o^{-1})_{dc}(i\omega_n)~ ~ ~ ~ ~ & (G_o^{-1})_{dd}(i\omega_n)
                      \end{array}
\right)~-~
\left( \begin{array}{ll}
        G_{cc}(i\omega_n)~ ~ ~ ~ ~ & G_{cd}(i\omega_n)\\
        G_{dc}(i\omega_n)~ ~ ~ ~ ~ & G_{dd}(i\omega_n)
                      \end{array}
\right)^{-1}
\label{mnfl4}
\end{eqnarray}

This leads to, for $N>1$, the self-energies with the following form at
low energies,

\begin{eqnarray}
&&\Sigma_{cc} (\omega) \sim
(\omega)^{\gamma_{cc}}\nonumber\\
&&\Sigma_{dd} (\omega) \sim
(\omega)^{\gamma_{dd}}\nonumber\\
&&\Sigma_{dc} (\omega) \sim
(\omega)^{\gamma_{dc}}
\label{mnfl8}
\end{eqnarray}
Here the exponents $\gamma_{cc}=2\epsilon_t^*-2$,
$\gamma_{dd}=1-\alpha$, and $\gamma_{dc}=2\epsilon_t^*+\beta-1$, where
$\alpha$ and $\beta$ are the exponents associated with $G_{dd}$ and
$G_{dc}$ given in Eq. (\ref{corr11}).

The form of the self-energies given in Eq. (\ref{mnfl8}) destroys
the quasiparticle pole. Therefore, the quasiparticle residue
vanishes; all the single particle excitations are incoherent.

We now turn to various physical correlation functions. Since both the
impurity spin and the charge couplings are irrelevant, we expect that
the multiparticle correlation functions in both impurity spin and
charge channels show power law behavior.
The spin and charge correlation functions are generally renormalized
differently.  The precise values of the exponents can be derived
using RG as discussed in Appendix \ref{sec:corr}. For example,

\begin{eqnarray}
&&<d_{\sigma}^\dagger d_{\sigma'}(\tau)d_{\sigma'}^\dagger d_{\sigma}(0)>
\sim (\tau)^{-\alpha_1}\nonumber\\
&&<d_{\sigma}^\dagger c_{\sigma'}(\tau)c_{\sigma'}^\dagger d_{\sigma}(0)>
\sim (\tau)^{-\alpha_2}\nonumber\\
&&<d_{\sigma}^\dagger c_{\sigma'}^\dagger (\tau)c_{\sigma'}d_{\sigma}(0)>
\sim (\tau)^{-\alpha_3}
\label{twopar}
\end{eqnarray}
where to leading order
\begin{eqnarray}
\alpha_1=&&2\epsilon_j^*(1-\delta_{\sigma,\sigma'})\nonumber\\
\alpha_2=&&2\epsilon_t^*\delta_{\sigma,\sigma'}
+2\epsilon_{tj}^*(1-\delta_{\sigma,\sigma'})\nonumber\\
\alpha_3=&&2\epsilon_t^*(-{\delta_1^* \over \pi},-{\delta_2^* \over
\pi})
\delta_{\sigma,\sigma'}+2\epsilon_{tj}^*
(-{\delta_1^* \over \pi},-{\delta_2^* \over \pi})
(1-\delta_{\sigma,\sigma'})
\label{twopar.exp}
\end{eqnarray}
in which $\epsilon_{tj}^*({\delta_1^* \over \pi},{\delta_2^* \over
\pi})
=((1-{\delta_1^* \over
{\pi}})^2+({\delta_1^* \over \pi}+{\delta_2^* \over {\pi}})^2+
(N-2)({\delta_1^* \over {\pi}})^2)/2$.

These exponents correspond to a divergent $d^\dagger c^\dagger$
pairing susceptibility and vanishing excitonic and $d$-electron
correlation functions at low energies. The divergent pairing
susceptibility signals a superconducting instability at a finite
temperature ${\rm T_c}$, with an anomalous normal state described by the
non-Fermi liquid state. We will comment on this further in section
\ref{sec:conclu}.

We now turn to the intermediate coupling phase. A critical chemical
potential $\mu_c$ also exists due to the irrelevant hybridization, as
was established in Section \ref{sec:flow5}. It again corresponds to a
{\it finite range} of densities in the lattice model, at which the
mixed valence behavior persists to zero temperature.

Since the exchange coupling is relevant, a resonance forms in the spin
channel below the corresponding coherence energy. Therefore, spin
excitations form quasiparticles which are expected to have
Fermi-surface features. The local spin susceptibility is regular. On
the other hand, the local charge susceptibility retains the algebraic
behavior. The excitonic correlation function still vanishes in the zero
frequency limit, while the superconducting  $d^\dagger c^\dagger$
susceptibility still diverges at low frequencies algebraically: the
system is again expected to be a superconductor with an anomalous
normal state.

To summarize, the non-Fermi liquid phases characterize incoherent
metals with vanishing quasiparticle residue, self-similar
local correlation functions, and asymptotically decoupled charge and
spin excitations. In addition, they occur over a range of electron
densities due to the pinning effect.

\subsection{Fermi Liquid Phase}
\label{sec:phases2}

We now turn to the region where the hybridization is relevant.
The system then flows away from the zero fugacity fixed points.
The new fixed point is beyond the reach of the perturbative RG. The
nature of the fixed point can be studied by other approaches
describing the strong coupling phases. One approach is the large
N limit of the model with fixed finite values of $V_1$ and
$V_2$.\cite{slaveboson,Grilli}

In the large N mean field theory applied directly to the lattice model
(\ref{hamiltonian.ehm}),
the renormalized hybridization never vanishes, and the
density is a continuous function of the chemical potential. The
$d$-electron spectral function is smeared into a coherent
Abrikosov-Suhl resonance.
Low energy excitations are coherent, and are
composed of conduction electrons and heavy electrons with finite
hybridization. The system is a Fermi-liquid.

That the large N mean field theory forces the system to be a Fermi liquid can
be
easily understood from our scaling equations (\ref{scaling}). When the
large N limit is taken, with a fixed finite values of $V_1$ and $V_2$,
the associated phase shifts given in Eqs. (\ref{phaseshift1}) and
(\ref{phaseshift12}) vanish.
Hence, the anomalous dimensions of the spin and charge flippings,
defined in  Eqs. (\ref{stiffness3}) and (\ref{phaseshift1}), fall in
the basin of attraction of the strong coupling fixed points.
Physically,  the excitonic and orthogonality contributions
are suppressed by a factor of ${1 \over N}$.

\subsection{Qualitative Phase Diagram of the Extended Hubbard model}
\label{sec:phases3}

The qualitative phase diagram, in terms of the interaction strength
and electron density, is summarized in Fig. \ref{phase}.

Fig. \ref{phase}(a) corresponds to varying the interaction strength
(in the $\epsilon_t-\epsilon_j$ plane in Fig. \ref{pdeps}), as that
characterized in Fig. \ref{criti}(a). For interaction strength such
that $\epsilon_1$ is above the threshold value $\epsilon_c$ given in
Fig. \ref{criti}(a), and over a range of electron density given by the
solid-line shaded region, a mixed valence metallic non-Fermi liquid
state occurs down to lowest temperatures (see, however, the discussion
about long range ordering in Section \ref{sec:conclu}). This range of density
corresponds to the critical chemical potential $\mu_c$ in Figs.
\ref{crossover}(b) and \ref{nmu}(b). For electron densities outside
this range, within the dashed-line shaded region, non-Fermi liquid
behavior occurs above a crossover temperature, shown
in Fig. \ref{crossover}(b).  The non-Fermi liquid state is
characterized by incoherent charge and spin excitations, which are
asymptotically decoupled.

For interaction strength such that $\epsilon_1$ is below  the
threshold value $\epsilon_c$, the system is a Fermi liquid. It
corresponds to the strong coupling mixed valence state of the
associated generalized Anderson model. Near the threshold interaction
strength, the Fermi energy is given by Eq. (\ref{critical}) with the
exponent $\eta={1 \over 2}$.

In Fig. \ref{phase}(b), we vary the interaction strength as that
characterized in Fig. \ref{criti}(b). For interaction strength such
that $\epsilon_2$ is above the threshold value $\epsilon_{c2}$ in Fig.
\ref{criti}(b),  a (weak coupling mixed valence) metallic non-Fermi
liquid phase again occurs down to zero temperature over a range of
electron density within the solid-horizontal-line shaded region, and
above a crossover temperature for electron densities within the
dashed-horizontal-line shaded region. The non-Fermi liquid state is
the same as that in Fig. \ref{phase}(a).

For interaction strength such that $\epsilon_2$ is between
$\epsilon_{c1}$ and $\epsilon_{c2}$ of
Fig. \ref{criti}(b), and over a range of electron density given by
the solid-vertical-line shaded region, an (intermediate coupling mixed
valence) metallic non-Fermi liquid phase occurs down to lowest
temperatures. For electron densities within the dashed-vertical-line
shaded region, the non-Fermi liquid behavior occurs above a crossover
temperature given in Fig. \ref{crossover}(b). The non-Fermi liquid
state is characterized by coherent spin excitations and incoherent
charge excitations, which also decouple asymptotically at low
energies. The spin coherence energy near $\epsilon_{c1}$ is given by
Eq. (\ref{critical}) with the exponent $\eta={1}$.

Finally, for interaction strength such that $\epsilon_2$ is below
the threshold value $\epsilon_{c1}$, the system is a Fermi liquid,
corresponding to the strong coupling state.

\subsection{Dependence on the Form of the Density of States}
\label{sec:phases4}

Thus far, we have focused on the extended Hubbard model with a
Lorentzian density of states. The non-Fermi liquid phases occur
due to the persistence of asymptotically degenerate states at the
Fermi level. This occurs because the quantum hopping amplitude (the
hybridization) which mixes these degenerate states is irrelevant.
These couplings are irrelevant because of the slow relaxation of the
electron bath to the hopping
between different local charge states. Such a large reaction exists
whenever there are non-zero density of states of the electron bath at
the Fermi level. In the Lorentzian case, there is always a finite
density of states for the bath electron.

To establish the generality of our results, we consider the extended
Hubbard model defined on the Bethe lattice with infinite connectivity.
In this case, the bare density of states is bounded.\cite{vanDogen,Rozenberg}
The Weiss field $G_o^{-1}$, which is a $2 \times 2$ matrix in our
case, can be explicitly related to the full Green's function,

\begin{eqnarray}
{G_o}^{-1}(i\omega_n) =
{g_o}^{-1}(i\omega_n) -\sum_{ij}t_{oi}G_{ij}t_{jo}
\label{bethe}
\end{eqnarray}
where ${g_o}^{-1}$ is the inverse propogator of the corresponding
non-interacting problem. More explicitly,

\begin{eqnarray}
&&(G_o^{-1})_{dd}=i\omega_n+\mu-E_d^o\nonumber\\
&&(G_o^{-1})_{cd}=t\nonumber\\
&&(G_o^{-1})_{cc} = i\omega_n + \mu -(t_{cc})^2G_{cc}
\label{bethe2}
\end{eqnarray}

Such a relation demonstrates that, when the renormalized $c$-electron
density of states is non zero at the chemical potential, the Weiss
field $(G_o^{-1})_{cc}$ will have a non-zero imaginary part. When
the hybridization $t$ is small, the $c$-electron acts as the electron
bath that the local $d-$electron degrees of freedom couple to.  Sine
the $c-$electron Green's function does not show infrared divergences,
as is discussed in Section \ref{sec:flow3} and \ref{sec:corr}, the
local Green's function of the electron bath has the long time
asymptotic ${1 \over \tau}$ behavior. This establishes that,  a finite
density of states for the electron bath at the Fermi level is indeed
self-consistent. This finite density of states can be effectively
modeled as the inverse Lorentzian width.\cite{new}

Therefore, our results which are explicitly derived for the extended
Hubbard model with a Lorentzian density of states is expected to
reflect qualitatively the physics of the extended Hubbard model away
 from half-filling. At the quantitative level, one has to determine the
charge and spin stiffness constants $\epsilon_t$ and $\epsilon_j$,
used in constructing the phase diagram in Fig. \ref{pdeps}, through
integrating out non-universal high energy dynamics. The phase diagram
will be altered correspondingly.

Finally, within the model with zero direct $d-$electron
hopping, only
$(G_o^{-1})_{cc}$ has a non-trivial form. Hence, direct $d-$electron
hopping effects are not generated. How new energy scales might be
generated from an explicit direct $d-$electron hopping term
is an important issue to be further studied.

\bigskip
\section{Conclusion}
\label{sec:conclu}

We have carried out a detailed study of an extended Hubbard model in
infinite dimensions, where the model reduces to a generalized
asymmetric Anderson problem in a self-consistent medium.
Within the associated impurity problem, we carried out a systematic
renormalization group analysis, and identified three kinds of mixed
valence behavior. Each of these states is associated with a
distinctive flow of spin and charge couplings. The usual strong
coupling mixed valence state has spin and charge couplings both
relevant; The weak coupling mixed valence state has spin and charge
couplings both irrelevant; And the intermediate coupling mixed valence
state has a relevant spin coupling and irrelevant charge coupling.
The strong coupling state corresponds  to a Fermi liquid phase of the
extended Hubbard model, while both the weak coupling and the
intermediate coupling mixed valence states describe  metallic
non-Fermi liquid phases of the extended Hubbard model.

The non-Fermi liquid phases are metallic states with incoherent charge
excitations. Within these incoherent metallic states, the local charge
modes are necessarily close to the Fermi level for charge
transport to occur. Such a pinning effect makes the non-Fermi liquid
phases to occur over {\it a range of electron
densities}. Hence, no fine tuning of parameters is necessary.
The pinning of the chemical potential leads to an
infinite compressibility at zero temperature: there is a jump in the
$n$ vs. $\mu$ curve, as shown in Fig. \ref{nmu}(b). It is
interesting to note that, an infinite compressibility has been
demonstrated in the numerical studies of the Hubbard model in the
limit of small dopings away from half-filling.\cite{Imada}

The non-Fermi liquid phases have vanishing quasiparticle residue: all
excitations are incoherent. In addition, charge and spin excitations
decouple asymptotically at low energies. Finally, certain local
correlation functions show algebraic behavior, reflecting the
self-similar nature of these incoherent metallic states. These are the
local correlations in both the charge and spin channels in the weak
coupling phase, and in charge but not spin channel in the intermediate
coupling phase. These properties have striking analogies to the
Luttinger liquid in one dimension.\cite{Haldane1D,Solyom}

The novel phase that we identify, i.e. the intermediate coupling phase,
has coherent spin excitations and incoherent charge excitations.
In connection with the possible realization of such a phase in
strongly correlated electron systems, we note that the normal state of
the high ${\rm T_c}$ copper oxides exhibits low energy spin dynamics
which are well described by renormalized quasiparticles with a
Luttinger Fermi surface, but charge dynamics which appear to be more
anomalous.\cite{Si} We will discuss physical properties of our
intermediate phase in more detail elsewhere.

We stress that at the end one ought to be able to express our results,
derived using a local method, in a language that would make contact
with perturbative calculations in the interactions. When this is done
we may be able to identify some similarities between our results and
those derived from other methods. This will also allow us to extend
our results to models at finite dimensions. In this regard, the
Fermi liquid phase is described in terms of the boson ``condensed''
phase within the slave boson formalism: the hybridization
renormalization is determined by the the quasi-condensate of the slave
boson, and the $d$-level renormalization is given by the mean field
value of a lagrangian multiplier enforcing the no-double-occupancy
constraint.\cite{slaveboson}
Our intermediate coupling incoherent metallic state has
similarities with the boson non-condensed phase,\cite{gauge}
which also has coherent spin excitation (``spinon'') and incoherent charge
excitation (``holon''). From our local treatment,
the characteristics of an incoherent metallic state lies in its
self-similar nature, a problem without scale. In both treatments of
these incoherent metallic states, the local gauge invariance which is
broken by the coherent hopping is finally restored.

We now comment on several aspects of our results which are important
to address in order to apply our results to finite dimensions
as well as to more general models.

In general, the limit of infinite dimensions separates the single
particle and collective excitations. As a result, when the system
undergoes an ordering instability, the one particle property does not
show precursor effect; The feedback of the long range order over the
single particle properties is down by powers of ${1 \over d}$. Within
the incoherent metallic phases, there exist asymptotically degenerate
local configurations, which is manifested in the algebraic low energy
behavior in various local correlation functions. In finite dimensions,
the coupling of the long range order on the local physics modifies the
local effective action at temperatures below the onset temperature
${\rm T_c}$, and has the effect of cutting
off the power law singularities. In our incoherent metallic states,
the most divergent local susceptibility is a pairing susceptibility.
Therefore, the instability is expected to be in the superconducting
channel. Our incoherent metallic states are then expected to be
relevant at temperatures $T_c << T << \omega_c$ in the same sense that
the paramagnetic Mott insulating phase is well defined for $T_N
<< T << \epsilon_F$ ($T_N$ being the Neel temperature).

Within the extended Hubbard model in infinite dimensions with a
Lorentzian density of states, the weak coupling incoherent metallic
state occurs over a range of attractive density-density interactions
and ferromagnetic spin interactions, while the intermediate coupling
incoherent metallic state occurs over a range of attractive charge
interactions and antiferromagnetic spin interactions. We emphasize
that, a Lorentzian form for the density of states suppresses the high
energy dynamics, enforcing the system to be always in the scaling
regime. In the generic case, the electron bath will be in the
universal regime only within a narrow energy range around the
Fermi level. The high energy dynamics have to be integrated out before
the renormalization group results can apply. Our results here, derived
for an extended Hubbard model with a Lorentzian density of states,
give the classification of the universality classes for a general
$N+1$ local configuration problem. To determine the basin of
attraction of the various phases, one has to first integrate out the
high energy dynamics beyond the scaling regime. Such a procedure makes
it possible that realistic Hamiltonians with all repulsive atomic
interactions fall in the basin of attraction of the non-Fermi liquid
phases.\cite{eps} Within generic models, the basin of attraction of
non-trivial phases can be determined numerically.

We conclude with a note on the methodology implications of our
analysis. The general renormalization group scheme developed here opens a local
view for
strongly correlated electron systems. The universality classes of a
correlated electron system can be determined by a small number of
atomic configurations coupled to an electron bath, which can be
treated through the general atomic expansion plus the renormalization
group procedure we
outlined here. The specifications of the phase diagram, on the other
hand, involves integrating out non-universal high energy dynamics,
which falls in the domain of quantum chemistry.\cite{eps}

\acknowledgments

We thank J. L. Cardy, A. M. Finkelstein, A. Georges, T. Giamarchi, P.
Le Doussal, A. Ruckenstein and C. Varma for useful discussions. This
work was supported by the NSF under grant DMR 922-4000.

\newpage
\appendix{Bosonization and the Level Shift}
\label{sec:bos}

In this appendix, we summarize the bosonization
procedure\cite{Schotte,Solyom} relevant to our discussion. We also
discuss the shift in the atomic levels that should be incorporated in
carrying out the bosonization.

Consider the projected Hamiltonian given in Eq. (\ref{halpha}),

\begin{eqnarray}
H_{\alpha} = E_{\alpha}+\sum_{\sigma} V_{\alpha}^{\sigma}
c^\dagger_{\sigma} c_{\sigma} + H_c
\label{bos1}
\end{eqnarray}
where

\begin{eqnarray}
H_c=\sum_{k,\sigma} E_k  c^\dagger_{k \sigma} c_{k \sigma}
\label{bos2.1}
\end{eqnarray}

For the asymptotic behavior, as long as the density of states
of the conduction electron is finite, a bosonization scheme can be
used. For the purpose of studying an impurity problem, we need to
consider only the S-wave component of the conduction electrons. We
label the radial dimension by $x$, which is extended to the negative
hafl-axis.

The Tomonaga bosons are defined as follows,
\begin{eqnarray}
&&b_{q,\sigma}^\dagger = i \sqrt{2 \pi \over q L} \sum_k
c_{k+q,\sigma}^\dagger c_{k,\sigma}\nonumber\\
&&b_{q,\sigma} = -i \sqrt{2 \pi \over q L} \sum_k c_{k,\sigma}^\dagger
c_{k+q,\sigma}
\label{bos2.2}
\end{eqnarray}
where $q>0$, and $L$ is the size of the radial dimension.

The free electron Hamiltonian has the following form in terms of the
Tomonaga bosons,

\begin{eqnarray}
H_c = \sum_{q,\sigma} v_F q b_{q,\sigma}^\dagger b_{q,\sigma}
\label{bos3}
\end{eqnarray}
where the boson dispersion is determined by $v_F = 1/\rho_o$, the
inverse of the density of states at the Fermi level,

The conduction electron operator has the following form,

\begin{eqnarray}
c_{\sigma} (x) = {1\over \sqrt{2\pi a}} e^{-i \Phi_{\sigma}(x)}
\label{bos3.2}
\end{eqnarray}
where

\begin{eqnarray}
\Phi_{\sigma}(x) = \sum_q \sqrt{2\pi \over q L } ( b_q^\dagger
e^{-iqx} + b_q e^{iqx})
\label{bos4}
\end{eqnarray}

The electron density, $\rho_{\sigma} (x)= c_{\sigma}^\dagger
c_{\sigma}$, is related to the $\Phi_{\sigma}$ field as follows,

\begin{eqnarray}
\rho_{\sigma} (x) = {d \Phi_{\sigma} \over d x}
\label{bos5}
\end{eqnarray}

Therefore, the potential term can be represented as follows,

\begin{eqnarray}
V_{\alpha}^{\sigma} c_{\sigma}^\dagger
c_{\sigma}={\delta_{\alpha}^{\sigma} \over {\pi \rho_o}}
({d \Phi_{\sigma} \over d x})_{x=0}
\label{bos6}
\end{eqnarray}

We use the prescription that, all quantities in the bosonization
representation corresponds to the normal ordered quantities. In so
doing, a Hartree-like shift in energy has to be incorporated
separately. The exact form of such a level shift was discussed in
detail for the spinless case in Ref. \cite{fk}. It can be easily
generalized to arbitrary spin degeneracy N, leading to,

\begin{eqnarray}
\Delta E_{\alpha} = {1\over \beta} \sum_{\sigma} \sum_n
\{ln(g_0^{-1})-ln(g_0^{-1}-V_{\alpha}^{\sigma})\}e^{i\omega_n 0^\dagger}
\label{shift.short2}
\end{eqnarray}
where $g_o (i \omega_n) = \sum_k {1 \over i\omega_n +\mu -
\epsilon_k}$. The linear in V term in the above shift $\Delta
E_{\alpha}$ is given by the $q=0$ component of the Tomonaga boson.

Such a shift can be captured by expressing $H_{\alpha}$ in the
bosonization form,

\begin{eqnarray}
H_{\alpha} = H_c+ {E_{\alpha}}'+ \sum_{\gamma}
{\delta_{\alpha}^{\gamma} \over \pi
\rho_o} ( {d \Phi_{\gamma} \over d x})_{x=0}
\label{boson.2}
\end{eqnarray}
where $E_{\alpha}'= E_{\alpha} + \Delta E_{\alpha}$.

\bigskip
\appendix{Renormalization Group Equations}
\label{sec:scaling.h}

In this appendix, we present the details of the derivation for the
RG equations, which will be written in the general
form, for $M$ arbitrary local configurations $|\alpha>$.

The RG equations describe the flow of the dimensionless couplings as
the bandwidth $1 /\xi$ is reduced. The RG charges
are the fugacities $y_{\alpha\beta}$, the stiffness constants
$-K(\alpha, \beta )$, and the symmetry breaking fields $h_{\alpha}$.
They satisfy $y_{\alpha\alpha}=0$, $K (\alpha,\alpha)=0$, and
$\sum_{\alpha} h_{\alpha}=0$. These relations are preserved in the
renormalization process.

Integrating out the degrees of freedom in the range $(\xi, \xi+d\xi)$
leads to three kinds of renormalization contributions. One
contribution arises from rescaling the old cutoff $\xi$ by the new one
$\xi'=\xi+d\xi$ in Eqs. (\ref{sumoverhis}-\ref{hisaction2}). This
amounts to a renormalization of the fugacities and the fields,

\begin{eqnarray}
y_{\alpha_{i+1},\alpha_{i}} ~&&\rightarrow~
y_{\alpha_{i+1},\alpha_{i}}
(1 +  (d ln \xi) (1+K_{\alpha_{i+1},\alpha_{i}})) \nonumber\\
h_{\alpha_{i}} ~&&\rightarrow~h_{\alpha_{i}} (1+ d ln \xi )
\label{ren1}
\end{eqnarray}

The other contributions arise from the change in the integration
range in the partition function, Eq. (\ref{sumoverhis}).
The dominant renormalization arises from integrating out kink-pairs
whose separation falls in the range $(\xi, \xi+d\xi)$. There are two
kinds of contributions in this category.

The contributions from integrating out non-neutral pairs,  i.e. pairs
with the final state of the second kink different from the initial
state of the first kink, amounts to a change in the fugacities:

\begin{eqnarray}
y_{\alpha_{i+1},\alpha_{i}} ~\rightarrow~
y_{\alpha_{i+1},\alpha_{i}} + (d ln \xi) \sum_{\gamma}
y_{\alpha_{i+1},\gamma} y_{\gamma, \alpha_{i}}
e^{(h_{\alpha_{i+1}}+h_{\alpha_{i}}-2h_{\gamma})/2}
\label{ren2}
\end{eqnarray}
for $\alpha_{i+1} \ne \alpha_{i}$.

On the other hand, integrating out neutral pairs leads to a
renormalization of the stiffness constants,

\begin{eqnarray}
K_{\alpha_{i+1},\alpha_{i}} ~\rightarrow~
K_{\alpha_{i+1},\alpha_{i}}
&&- d( ln \xi ) \sum_{\gamma}
y^2_{\alpha_{i+1},\gamma} e^{h_{\alpha_{i+1}}-h_{\gamma}}
(K_{\alpha_{i+1},\alpha_{i}} +K_{\alpha_{i+1},\gamma}
-K_{\alpha_{i},\gamma})\nonumber\\
&&- d( ln \xi ) \sum_{\gamma}
y^2_{\alpha_{i},\gamma} e^{h_{\alpha_{i}}-h_{\gamma}}
(K_{\alpha_{i+1},\alpha_{i}}
+K_{\alpha_{i},\gamma}
-K_{\alpha_{i+1},\gamma})
\label{ren3}
\end{eqnarray}
as well as a renormalization in the field and in the free energy,

\begin{eqnarray}
h_{\alpha_{i}} ~&&\rightarrow~
h_{\alpha_{i}}
-d ( ln \xi) (\sum_{\gamma} y_{\alpha_i,\gamma}^2
e^{h_{\alpha_i}-h_{\gamma}} - {1 \over M}
\sum_{\gamma_1,\gamma_2}y_{\gamma_1,\gamma_2}^2
e^{h_{\gamma_1}-h_{\gamma_2}})\nonumber\\
F\xi ~&&\rightarrow~ F\xi
-d(ln\xi) {1 \over M} \sum_{\gamma_1,\gamma_2}y_{\gamma_1,\gamma_2}^2
e^{h_{\gamma_1}-h_{\gamma_2}}
\label{ren4}
\end{eqnarray}
Here the separation into the field renormalization and free energy
renormalization is unambiguously specified by requiring  that
$\sum_{\alpha} h_{\alpha} =0$ be preserved in the process of
renormalization.

Collecting all these terms, we derive the following general form of
the RG equations.

\begin{eqnarray}
{d y_{\alpha,\beta} \over {d ln \xi}} = &&(1+K(\alpha,\beta)) y_{\alpha,\beta}
+\sum_{\gamma} y_{\alpha,\gamma} y_{\gamma, \beta}
e^{(h_{\alpha}+h_{\beta}-2h_{\gamma})/2}\nonumber\\
{ d K(\alpha,\beta) \over d ln \xi } = &&-\sum_{\gamma}
y_{\alpha,\gamma}^2 e^{h_{\alpha}-h_{\gamma}}
(K(\alpha,\beta)+K(\alpha, \gamma) - K ( \beta,\gamma))\nonumber\\
&&-\sum_{\gamma} y_{\beta,\gamma}^2 e^{h_{\beta}-h_{\gamma}}
(K(\alpha,\beta)+K(\beta, \gamma) - K (\alpha,\gamma))\nonumber\\
{ d h_{\alpha} \over {d ln \xi}}= &&h_{\alpha}-\sum_{\gamma}
y_{\alpha,\gamma}^2 e^{h_{\alpha}-h_{\gamma}} +{1 \over M}
\sum_{\alpha,\beta}y_{\alpha,\beta}^2 e^{h_{\alpha}-h_{\beta}}
\nonumber\\
{ d {F\xi} \over {d ln \xi}}= &&F \xi -{1 \over M}
\sum_{\alpha,\beta}y_{\alpha,\beta}^2 e^{h_{\alpha}-h_{\beta}}
\label{scaling.22}
\end{eqnarray}

\bigskip
\appendix{Renormalization of Correlation Functions\\
in the Weak Coupling Regime}
\label{sec:corr}

In this appendix, we give the details of the derivation for the
renormalization of the exponents of the correlation functions
in the weak coupling regime.

The Green's function
\begin{eqnarray}
G_{\psi\psi}(\tau) = - <T \psi (\tau) \psi^\dagger(0)>
\label{corr1}
\end{eqnarray}
where $\psi^\dagger=(d^\dagger,c^\dagger)$, can be expanded in terms
of the flipping part of the Hamiltonian $H_f$,

\begin{eqnarray}
G_{\psi\psi}(\tau) &&=- {1 \over Z}  \int Dc~D d ~ ~T[\psi (\tau)
\psi^\dagger(0) exp [-(S_0 + \int_0^{\beta}d\tau H_f(\tau) )]
]\nonumber\\
&&={1\over Z}\sum_{n=0}^{\infty}~\int_{\xi_0}^{\beta-\xi_0}d\tau_n
...\int_{\xi_0}^{\tau_{i+1}-\xi_0} d \tau_i
...\int_{\xi_0}^{\tau_{2}-\xi_0} d \tau_1
 ~A(\alpha; \tau;0;\tau_n,...,\tau_1)
\label{corr1.2}
\end{eqnarray}
where the transition amplitude

\begin{eqnarray}
A(\alpha; \tau;0;\tau_n,...,\tau_1) = &&(-1)^n
\int Dc Dd \nonumber\\
&&<\alpha|T[ exp(-\beta H_0 ) \psi (\tau)
H_f(\tau_n)...H_f(\tau_i)...H_f(\tau_1) \psi^\dagger(0)] |\alpha>
\label{corr2}
\end{eqnarray}

Without loss of generality, we take $\tau>0$. For a given history
similar to that illustrated in Fig. \ref{hop}, the external time
$\tau$ lie between two flips, say $\tau_{k-1}<\tau<\tau_k$. The
transition amplitude has the following form once a complete set of
local states is inserted at every discrete time

\begin{eqnarray}
A(\alpha; \tau;0;&&\tau_n,...,\tau_1) = (-1)^n \sum_{\alpha_1, ...,
\alpha_n} \int Dc ~ exp[-H_{\alpha} (\beta-\tau_{n})]
Q(\alpha,\alpha_{n})...\times\nonumber\\
&&\times exp [-H_{\alpha_{k+1}}
(\tau_{k+1}-\tau_{k})]~Q(\alpha_{k+1},\alpha_{m})  exp[-H_{\alpha_{m}}
(\tau_{k}-\tau)]\times \nonumber\\
&&\times \psi~ ~exp [-H_{\alpha_{m'}} (\tau-\tau_{k-1})]
Q(\alpha_{m'},\alpha_{k-1})~ exp [-H_{\alpha_{k-1}}
(\tau_{k-1}-\tau_{k-2}) ]...\times \nonumber\\
&&\times exp [-H_{\alpha_2} (\tau_2-\tau_1)]~ ~Q(\alpha_2,\alpha_1)
 ~exp [-H_{\alpha_1}\tau_1 ]~\psi^\dagger
\label{corr3}
\end{eqnarray}
where $|\alpha_m>$ and $|\alpha_{m'}>$ label states after and before
the external time $\tau$. The operator $\psi^\dagger$, on the other
hand, flips from the local state $|\alpha>$ to $|\alpha_1>$. We
emphasize that, the external fermion operators place constraints on
the hopping history of the impurity: the insertion of a $d-$electron
creats a charge kink, while the insertion of a local $c-$electron does
not creat any kink.

Tracing out the conduction electron degrees of freedom,

\begin{eqnarray}
A(\alpha; \tau;&&0;\tau_n,...,\tau_1)= Z_c
\sum_{\alpha_{n+1}=\alpha,\alpha_1, ..., \alpha_n}
y'_{\alpha_{n+1}, \alpha_n}... y'_{\alpha_{k+1}, \alpha_m }
y'_{\alpha_{m'}, \alpha_{k-1} }...
y'_{\alpha_{2}, \alpha_1 }\times\nonumber\\
&&\times exp[E_{\alpha}'\tau_n-E_{\alpha_1}'\tau_1-\sum_{i=2}^{n-1}
E_{\alpha_{i+1}}' (\tau_{i+1}-\tau_i)] <O(\alpha_{n+1}, \alpha_n )
(\tau_n)...O(\alpha_{k+1}, \alpha_m ) (\tau_k)\times \nonumber\\
&&\times \Psi_{\alpha_m,\alpha_{m'}} (\tau) O(\alpha_{m'}, \alpha_{k-1} )
(\tau_{k-1})... O(\alpha_2, \alpha_1 ) (\tau_1)
\Psi_{\alpha_1,\alpha}^\dagger(0)>
\label{corr4}
\end{eqnarray}
where $\Psi_{\alpha,\beta}^\dagger =(D_{\alpha,\beta}^\dagger,
C_{\alpha,\beta}^\dagger)$ have the following form,

\begin{eqnarray}
&&(D_{\sigma})_{\alpha,\beta}(\tau) \equiv exp (ie_D^{\gamma}
\Phi_{\gamma}(\tau)) \delta_{\alpha,|0>} \delta_{\beta,|\sigma>}\nonumber\\
&&(C_{\sigma})_{\alpha,\beta}(\tau) \equiv exp (ie_C^{\gamma}
\Phi_{\gamma}(\tau)) \delta_{\alpha,\beta}
\label{Dtau}
\end{eqnarray}
with,

\begin{eqnarray}
&&e_D^{\gamma}={\delta_1 \over \pi} + {\delta_2 \over \pi}
\delta_{\gamma,\sigma}\nonumber\\
&&e_C^{\gamma}=-\delta_{\gamma,\sigma}
\label{Dtau2}
\end{eqnarray}

The transition amplitude now has again a simple form in boson
representation through

\begin{eqnarray}
<O(\alpha_{n+1}, \alpha_n ) (\tau_n)... &&O(\alpha_{k+1}, \alpha_m )
(\tau_k) \Psi_{\sigma} (\tau) O(\alpha_{m'}, \alpha_{k-1} )
(\tau_{k-1})... O(\alpha_2, \alpha_1 ) (\tau_1)
\Psi_{\sigma}^\dagger(0)> \nonumber\\
=&& \int exp (-S_c+ \int_0^{\beta} D \tau'
\sum_{\gamma}j_{\gamma}(\tau') \Phi_{\gamma} (\tau') )
\label{corr5}
\end{eqnarray}
where

\begin{eqnarray}
j_{\gamma}(\tau')  = &&{\sum '}_{i =1}^n \delta(\tau'-\tau_i)
e_{\alpha_{i+1},\alpha_i}^{\gamma} +\delta (\tau' - \tau_{k-1})
e_{\alpha_{m'}, \alpha_{k-1}}^{\gamma} +\delta (\tau' - \tau_{k})
e_{\alpha_{k},\alpha_m}^{\gamma}\nonumber\\
&&+\delta (\tau' - \tau_1) e_{\alpha_1, \alpha_2}^{\gamma} +\delta
(\tau' - \tau_{n}) e_{\alpha_n,\alpha}^{\gamma} +\delta (\tau' - \tau)
e_{\Psi}^{\gamma} +\delta (\tau' ) e_{\Psi^\dagger}^{\gamma}
\label{corr6}
\end{eqnarray}
Here a primed sum means a summation which excludes ${n,1,k-1 , k}$.
Therefore,

\begin{eqnarray}
<O(\alpha_{n+1}, \alpha_n ) (\tau_n)... &&O(\alpha_{k+1}, \alpha_m )
(\tau_k) \Psi_{\sigma} (\tau) O(\alpha_{m'}, \alpha_{k-1} )
(\tau_{k-1})... O(\alpha_2, \alpha_1 ) (\tau_1)
\Psi_{\sigma}^\dagger(0)> \nonumber\\
=&& exp[-\sum_{\gamma} A^{\gamma}]
\label{corr7}
\end{eqnarray}
where

\begin{eqnarray}
A^{\gamma}= &&{1\over 2} {\sum_{i_1}}' {\sum_{i_2}}'
e_{\alpha_{i_1+1},\alpha_{i_1}}^{\gamma}
e_{\alpha_{i_2+1},\alpha_{i_2}}^{\gamma} \log {|\tau_{i_1}-\tau_{i_2}|
\over \xi_0} + {1\over 2} {\sum_{i_1}}'' {\sum_{i_2}}''
e_{\alpha_{i_1+1},\alpha_{i_1}}^{\gamma}
e_{\alpha_{i_2+1},\alpha_{i_2}}^{\gamma} \log {|\tau_{i_1}-\tau_{i_2}|
\over \xi_0} \nonumber\\
&&+{\sum_{i_1}}' {\sum_{i_2}}''
e_{\alpha_{i_1+1},\alpha_{i_1}}^{\gamma}
e_{\alpha_{i_2+1},\alpha_{i_2}}^{\gamma} \log {|\tau_{i_1}-\tau_{i_2}|
\over \xi_0} + {\sum_{i}}' e_{\alpha_{i+1},\alpha_{i}}^{\gamma}
e_{\Psi^\dagger}^{\gamma} \log {\tau_{i} \over \xi_0}\nonumber\\
&&+ {\sum_{i}}' e_{\alpha_{i+1},\alpha_{i}}^{\gamma}
e_{\Psi}^{\gamma} \log {|\tau_{i}-\tau | \over \xi_0}
+e_{\Psi^\dagger}^{\gamma} e_{\Psi}^{\gamma} \log {\tau \over
\xi_0}\nonumber\\
&&+{\sum_{i}}' h_i (\tau_{i+1}-\tau_i)
+h_m (\tau_{k}-\tau) +h_{m'} (\tau-\tau_{k-1})
+h_{\alpha} (\beta-\tau_n) + h_{\alpha_1} \tau_1
\label{corr7.2}
\end{eqnarray}
Here the double primed summation extends over the neighbors of the
$0$ and $\tau$, which includes $\tau_n, \tau_1, \tau_{k-1}, \tau_k$.

We are now ready to perform the RG calculation, by integrating out
degrees of freedom in the range $(\xi, \xi+d\xi)$.
This again leads to renormalization contributions of three kinds: a)
simple rescaling; b) integrating out non-neutral pairs; and c)
integrating neutral pairs. The contributions from simple rescaling of
the cutoff, as well as from integrating out the primed kink-pairs,
give rise to scaling of the fugacities, stiffness constants and the
fields, which are the same as those in the partition function.

On the other hand, integrating out the pairs formed between a doubly
primed kink and its neighbor in the primed kinks will contribute
directly to the renormalization of the correlation functions. For the
Green's functions, there are four terms contributing to the
renormalization of the propagators. Contributions from the pair of
kinks before and after the external time $\tau$ cancel unless the
original states before and after $\tau$ are different.

Collecting all these contributions, we arrive at the results

\begin{eqnarray}
{d (\alpha_{dd} ln(\tau /\xi )) \over d ln \xi} ~&&=~-2 \sum_m \sum_{\gamma}
( e_D^m e_{\alpha_m,\gamma}^m y_{\alpha_m,\gamma}^2
+ e_{D'}^m e_{\alpha_{m'},\gamma}^m y_{\alpha_{m'},\gamma}^2 )
ln {\tau \over \xi }\nonumber\\
{d (\alpha_{cd} ln(\tau /\xi )) \over d ln \xi} ~&&=~ \sum_m
\sum_{\lambda} \sum_{\gamma}  e_{\lambda,\lambda}^m (
e_{\alpha_m,\gamma}^m y_{\alpha_m,\gamma}^2 -e_{\alpha_{m'},\gamma}^m
y_{\alpha_{m'},\gamma}^2 ) ln {\tau \over \xi }\nonumber\\
{d (\alpha_{cc}ln(\tau /\xi )) \over d ln \xi} ~&&=~ 0
\label{corr9}
\end{eqnarray}
where $\alpha_{dd}$, $\alpha_{cc}$, $\alpha_{dc}$ are the exponents
for $G_{dd}$,$G_{cc}$,$G_{dc}$ respectively, in the algebraic
dependence on imaginary time. Integrating from $\xi=\xi_0$ and
$\xi=\tau$ gives rise to the correction to the Green's function due to
the fugacities.

The two particle correlation functions can be calculated following the
same procedure. The results are given in section \ref{sec:phases1}.

\figure{Hopping sequences in the atomic representation. Here
$\tau_i$, for $i=1,...,n$, labels the imaginary time at which the
atomic configurations hops from $|\alpha_i>$ to
$|\alpha_{i+1}>$.\label{hop}}

\figure{Phase diagram of the generalized Anderson model in the mixed
valence regime. Here $\epsilon_t$ and $\epsilon_j$ label the charge
and spin stiffness constants defined in the text. The vertical thick
line, the horizontal thick line, and the dashed line are the
boundaries between the strong couping, weak coupling, and intermediate
coupling mixed valence states.\label{pdeps}}

\figure{Zero temperature transition sequences when the interactions
are varied such that within the $\epsilon_t-\epsilon_j$ plane of Fig.
\ref{pdeps}, (a) a one stage transition takes place from
the weak coupling to the strong coupling state; and (b) two stage
transitions take place from the weak coupling, through the
intermediate coupling, into the strong coupling state. Here
$\epsilon_c$, $\epsilon_{c1}$ and $\epsilon_{c2}$ label points on the
vertical thick line, the horizontal thick line and the dashed line
respectively in Fig. \ref{pdeps}.\label{criti}}

\figure{Crossover diagram in terms of temperature ($T$) and
chemical potential ($\mu$) (a) for the strong coupling case. The dashed
lines correspond to the crossovers, and $\Delta^*$ is the renormalized
resonance width; (b) for both the weak coupling and the intermediate
coupling cases. Here $\mu_c$ labels the critical
point.\label{crossover}}

\figure{Electron density ($n$) as a function of chemical potential
($\mu$) (a) for the strong coupling case and (b) for both the weak
coupling and the intermediate coupling cases.\label{nmu}}

\figure{Schematic interaction$-$density phase diagram for two ways of
varying the interactions. Here $\epsilon_1$ and $\epsilon_2$ are the
same as those in Fig. \ref{criti}. (a) For $\epsilon_1$ above
$\epsilon_c$, non-Fermi liquid phase with both charge and spin
excitations incoherent occurs till zero temperature within the
solid-line shaded density range, and above a crossover temperature
within the dashed-line shaded density range. Fermi liquid phase occurs
otherwise. (b) For $\epsilon_2$ above $\epsilon_{c1}$, non-Fermi
liquid phase of (a) occurs. For $\epsilon_2$ between $\epsilon_{c1}$
and $\epsilon_{c2}$, non-Fermi liquid phase with incoherent charge
excitations but coherent spin excitations occurs till zero temperature
within the solid-vertical-line shaded density range, and above a
crossover temperature within dashed-vertical-line shaded density
range. Fermi liquid phase occurs otherwise.\label{phase}}


\begin{references}
\bibitem[38]{finiteU} The finite U problem can be dealt with through
the scaling equations in a finite field given in Appendix
\ref{sec:scaling.h}.
\bibitem[40]{note3} The $(-1)^n$ factor is absorbed by the ordering of
the conduction electron operators in the corresponding correlation function.
\bibitem[41]{finiteT} At finite temperature $T={1 \over \beta}$,
$1/\tau$ is replaced by ${\pi \over \beta} / sin ({\pi \over
\beta}\tau )$.
\bibitem[42]{rotat-inv} Because of rotational invariance in spin space
the second and the fourth equations should be identical. This is
achieved explicitly by expanding to quadratic order in the small
quantities $y_j={(1-\epsilon_j) \over 2}$.
\end{references}
\end{document}